\documentclass[prx,reprint,superscriptaddress,tightenlines,twocolumn,showpacs,floatfix,longbibliography,nofootinbib]{revtex4-2}
\pdfoutput=1

\usepackage{latexsym,amsmath,amsfonts,amssymb,bm,empheq}   
\usepackage{nicefrac} 
\usepackage[dvipsnames]{xcolor}
\usepackage{booktabs}
\usepackage{tikz}   
\usetikzlibrary{arrows.meta,decorations.markings,bending}
\usepackage{subfigure}  

\definecolor{nblue}{RGB}{28,130,185}
\definecolor{cgreen}{RGB}{76,153,0}
\definecolor{myorange}{RGB}{245,156,74}

\usepackage{hyperref}
\hypersetup{
  colorlinks=true,
  citecolor=magenta,
  urlcolor=-myorange
}

\usepackage{multirow}

\usepackage{array}
\AtBeginDocument{
\cmidrulewidth=.03em
}

\newcommand{\charlie}[1]{{\color{black}{#1}}}

\begin{document}

\title{Nonequilibrium polarity-induced mechanism for chemotaxis:\\
emergent Galilean symmetry and exact scaling exponents}

\author{Saeed Mahdisoltani}
\thanks{These three authors contributed equally.}
\affiliation{Rudolf Peierls Centre for Theoretical Physics, University of Oxford, Oxford OX1 3PU, United Kingdom}
\affiliation{Max Planck Institute for Dynamics and Self-Organization (MPIDS), D-37077 G\"ottingen, Germany}

\author{Riccardo Ben Al\`{i} Zinati}
\thanks{These three authors contributed equally.}
\affiliation{SISSA --- International School for Advanced Studies \& INFN, via Bonomea 265, I-34136 Trieste, Italy}

\author{Charlie Duclut}
\thanks{These three authors contributed equally.}
\affiliation{Max-Planck-Institut f\"ur Physik komplexer Systeme, N\"othnitzer Str.~38, D-01187 Dresden, Germany}

\author{Andrea Gambassi}
\affiliation{SISSA --- International School for Advanced Studies \& INFN, via Bonomea 265, I-34136 Trieste, Italy}

\author{Ramin Golestanian}
\email{ramin.golestanian@ds.mpg.de}
\affiliation{Rudolf Peierls Centre for Theoretical Physics, University of Oxford, Oxford OX1 3PU, United Kingdom}
\affiliation{Max Planck Institute for Dynamics and Self-Organization (MPIDS), D-37077 G\"ottingen, Germany}

\date{\today}
\begin{abstract}

A generically observed mechanism that drives the self-organization of living systems is interaction via chemical signals among the individual elements---which may represent cells, bacteria, or even enzymes. 
Here we  {propose} a novel mechanism for such interactions, in the context of chemotaxis, which originates from the polarity of the particles and  {which} generalizes the well-known Keller--Segel interaction term. 
We study the  {resulting} large-scale dynamical properties of a system of such 
chemotactic particles using the exact stochastic formulation of Dean  {and} Kawasaki along with dynamical renormalization group analysis of the critical state of the system. 
At this critical point, an \textit{emergent} ``Galilean'' symmetry is identified, 
which allows us to obtain the dynamical scaling exponents exactly; these exponents reveal superdiffusive density fluctuations and non-Poissonian number fluctuations. 
We expect our results to shed light on how molecular regulation of chemotactic circuits can determine large-scale behavior of cell colonies and tissues.

\end{abstract}

\maketitle

\section{Introduction}	\label{sec:intro}

Characterizing the emergence of macroscopic properties in colonies of prokaryotic \cite{adler1966,ben-jacob2000} and eukaryotic \cite{levine2013} cells based on the complicated chemical interactions among the individuals in the colony is a long-standing endeavor in various areas of biology such as morphogenesis \cite{hogan1999,crick1970,friedl2009}, tissue growth and homeostasis \cite{tzur2009}, wound healing \cite{schneider2010}, and cancer metastasis \cite{hanahan2011,bockhorn2007}. 
\color{black}
A prevalent interaction in such contexts is \textit{chemotaxis}:
\color{black}
the ability of bacteria and cells to detect the changes in the concentrations of specific chemical molecules in their surrounding media \cite{iijima2002} and to respond to them by adjusting their 
polarity or direction of motion \cite{roussos2011,iglesias2008,alon1999}.
Although the detailed mechanisms responsible for chemotaxis in cells are rather complex \cite{barkai1997,sourjik2004,wadhams2004,tu2008,emonet2008,tu2013,westendorf2013}, the phenomenon seems to emerge generically in nature. Moreover, it has also been observed in smaller and more primitive systems such as enzymes \cite{dey14,agudo2018,jee19} and synthetic catalytically active colloids \cite{Kapral2012,Soto2014,yan2016,illien2017,stark2018,speck2017,Colberg2017}.
Chemotactic interactions are typically long-range as the transmitting molecules decay very slowly and, therefore, it is not surprising that these interactions share some of the features of other long-range interactions such as the electrostatic and gravitational ones. For instance, it has been shown that the self-organization of chemotactic species resembles the formation of galaxies in astrophysics, as well as the large-scale vortices in two-dimensional turbulence \cite{chavanis2010, chavanis2008}.

Chemotactic systems are often studied through various formulations of the Keller--Segel (KS) model \cite{keller1970,keller1971,hillen2009} which are phenomenological mean-field approximations that model chemotaxis as a directed motion guided by chemical gradients \cite{keller1971,tsori2004, newman2004}.
These models and their stochastic variations \cite{grima2005,golestanian2009,sengupta2009,taktikos2012} have proven useful in studying the chemotactic collapse of bacteria \cite{brenner1998,chavanis2004} and collective behavior of active colloids \cite{golestanian2012,cohen2014}. 
Other generalizations of these models, which incorporate the polarity of the active particles and their active alignment, have been used to study collective properties of synthetic active Janus particles \cite{saha2014,liebchen2017} as well as chemotaxis for trail-following bacteria \cite{kranz2016,gelimson2016}. 

Owing to the large number of degrees of freedom usually involved in a colony, coarse-grained descriptions are particularly useful in studying these and more general active systems \cite{marchetti2013,Gompper2020}. In cases where the correlations are long-range and collective phenomena emerge, standard field-theoretical approaches have been applied to a wide range of models of biological or synthetic colonies such as flocks of birds, schools of fish, aggregations of molecular motors, and dividing chemotactic particles \cite{vicsek1995, toner1995, toner2012, risler2004, gelimson2015}.
Similar approaches have been used to study nonequilibrium field theories 
with applications to active phase separation and motility-induced phase separation \cite{cates2015,wittkowski2014,caballero2018,caballero2018a,soto2014b}.

In the present work, we investigate the macroscopic properties of a collection of particles with generalized chemotactic response taking into account both the KS response and the 
 {polarity} of the particles induced by changes in the chemical field. 
We discuss the  {possible} microscopic origin of this 
 {polarity-induced} chemotaxis in a biological context, which is  {based on a} simple toy model and rigorous derivations using mean-field equations. 
Aiming to focus on the critical state of the system, where the fluctuations are most relevant as  {their spatial} correlation length diverges, we employ the Dean--Kawasaki (DK) approach \cite{dean1996,kawasaki1994} to account for the statistical correlations that were neglected in the mean-field KS equation, by introducing noise in our description. 
The resulting stochastic field equation that governs the particle density  {predicts the existence of} a dispersed phase of the system and a collapsed one, separated by a critical state. 
The critical dynamics is then analyzed using a dynamical renormalization group (RG) treatment~\cite{forster1977,medina1989,tauber2014} to obtain the emergent macroscopic properties of the chemotactic system based on the interactions between its individuals. 

In these stochastic field equations, we identify an emergent symmetry, which turns out to coincide with the ``Galilean'' symmetry known in the apparently unrelated context of the Kardar--Parisi--Zhang (KPZ) equation~\cite{kardar1986,medina1989,frey1994}. 
This symmetry is realized
when the diffusion of the chemical signals is considerably faster than that of the particles,  {such that} the chemical signals emitted by particles can diffuse  {efficiently} across the  {entire} system before the particle themselves displace significantly or the chemicals possibly degrade, effectively establishing long-range interactions among the constituents.
This symmetry, which we expect to be realized in a wide class of  long-range interacting systems,  including gravitational and electrostatic settings, boosts our understanding of the formal structure of the field-theoretical description of stochastic dynamics of the system by providing nonperturbative relationships among correlation functions (i.e.,~Ward identities) that yield \textit{exact} scaling exponents.

The analysis of the scaling properties of the stochastic evolution equation is performed both with a nonconserved noise, relevant in the case where the number of particles is conserved only on average, and with a conserved noise. In both cases, the exact exponents we obtain predict superdiffusion at the critical state, while the  {magnitude} of the fluctuations  {of} the particle number depends on the nature of the noise: a conserved noise suppresses these fluctuations and the distribution becomes hyperuniform, whereas a nonconserved noise enhances the fluctuations and leads to giant number fluctuations.
These scaling properties are observed at the stable one-loop RG fixed points.

This work highlights the crucial role of the polarity-induced chemotactic mechanisms which are often overlooked in theoretical models. 
Although these may stem from subleading contributions at the level of the single isolated particle, we show that one such term becomes as relevant as the KS term in the presence of many particles. 
Moreover, we show that this 
polarity-induced nonlinearity is a purely nonequilibrium interaction, which indicates that the system does not reach an equilibrium state in the long-time limit. 
It therefore stands in contrast with the traditional KS chemotactic drift which is essentially an equilibrium-like interaction, as it is derived from a potential~\cite{chavanis2010}. 
%


The rest of the paper is organized as follows:  
In Sec.~\ref{sec:polareffects}, we first provide a mechanistic view of the biological context of our work and the phenomenology arising from the 
 {polarity-induced} chemotactic interactions we consider. 

In Sec.~\ref{sec:stoch} we use the DK approach to obtain the exact equation governing the instantaneous particle density. 
This is then phenomenologically extended to account for the case where the chemotactic response of the particles can be activated or inactivated by adding a linear growth term to the DK equation. 
By expanding the DK equation around a state with uniform density, we then obtain the Langevin equation of the fluctuating particle density at this mesoscopic scale, and we highlight its Galilean invariance. 
In Sec.~\ref{sec:RG} the scaling behavior of  {this} Langevin  {equation} is examined, and is supported by the RG calculations.
We present the features of the resulting RG flows in Sec.~\ref{sec:results} and then discuss the scaling exponents, which are obtained exactly due to the Galilean symmetry. 
In Sec.~\ref{sec:symmetry}, we discuss the emergence of the Galilean symmetry and its implication in this work. 
Finally, we present the conclusions and outlook of this work in Sec.~\ref{sec:conc}.
There are eight appendices that contain additional information regarding two toy models in which the 
 {effect of the polarity} is illustrated (Appendix \ref{app_toymodel}),
the most relevant aspects of the moment expansion supporting the results discussed in Sec.~\ref{sec:polareffects} (Appendix~\ref{app:moment}), 
the validity of detailed balance in our chemotactic field theory (Appendix~\ref{app_detailedbalance}),
the gradient expansion and power counting (Appendix~\ref{app_powerCounting}),
the details of the RG calculations (Appendices~\ref{app_rgprocedure} and~\ref{app_rgnon}),
the analysis of the RG flows in various spatial dimensions (Appendix~\ref{app_rgflowsdim}),
 and the thorough discussion of the moment expansion, anticipated in Appendix~\ref{app:moment}, for a more general chemotactic model including self-propulsion and nematic alignment of the particles (Appendix~\ref{app:moment_long}).

%

 \section{Polarity effects in generalized chemotaxis} 	\label{sec:polareffects}

In this section, we present the phenomenology of the 
 {polarity} effects in chemotaxis within a biological context, which leads to a generalization of the KS equation of motion for a Brownian particle in Eq.~\eqref{eq:cell_velocity}.


The chemotactic response of a {\em single} cell in a medium with a concentration field $\Phi(\mathbf{x},t)$ is commonly described by a drift velocity
%
$
\bm{v}_{\rm KS} = \nu_1 \nabla \Phi 
$
%
first introduced by Keller and Segel \cite{keller1970,keller1971}. This biased motion can be a result of temporal sensing mechanisms \cite{sourjik2004,wadhams2004}, as observed in prokaryotes such as \textit{E.~coli}, or spatial sensing \cite{iglesias2008,levine2013,westendorf2013}, as observed in eukaryotes. 

We propose an independent mechanism by which the cell  {polarity} can influence  {the} chemotactic response. 
To illustrate this mechanism, we consider a chemotactic cell for which the local distribution of the chemical sensory units determines the feedback onto the motility machinery. 
For instance, it has been reported that the distribution of the chemical sensing units on the membrane of neutrophils changes with the chemical gradient in the surrounding \cite{servant2000}. 
In such cases, one can assign a unit vector $\mathbf{n}$ to each cell, henceforth called the \textit{cell polarity}, that characterizes the possible anisotropy in the response of the cell to the chemical gradients. 
This polarity can, for instance, result from asymmetries in the shape of the cells or in the distribution of the sensory units present on the surface of the cell.
The overall movement of the cell is then influenced by the instantaneous direction of the cell polarity, which contributes with
$
\bm{v}_{\rm p} = \nu_2 \mathbf{n} \cdot \nabla \nabla \Phi.
$
to the local cell velocity.
Note that this expression may be seen as the chemotactic equivalent of the electrostatic force acting on a dipole in an external electric potential, generalizing the KS ``monopole'' expression. 
For illustration purposes, in Appendix \ref{app_toymodel} we present a minimal microscopic model based on which the emergence of 
these two effects can be explicitly derived by simple calculations.   
The resultant of these two local responses to chemical gradients, 
$\bm{v}_{\rm KS}$ and $\bm{v}_{\rm p}$, 
determines the drift velocity of the chemotactic particle as
\begin{equation} \label{eq:v-nu1-nu2-1}
\begin{split}
\frac{\mathrm{d}\mathbf{r}}{\mathrm{d}t} &= 
\bm{v}_{\rm KS} + \bm{v}_{\rm p} + \bm{\xi} \\
&= 
\nu_1 \nabla \Phi + 
\nu_2 \mathbf{n} \cdot \nabla \nabla \Phi +
\bm{\xi}, 
\end{split}
\end{equation}
where we have also included a noise term $\bm{\xi}$ 
with vanishing mean and variance $\left\langle \xi_i(t) \xi_j(t') \right\rangle = 2 D\delta_{ij} \delta(t-t')$ to account for the Brownian motion of the particle.
%

%
\begin{figure}[t]
\includegraphics[width=0.48\textwidth]{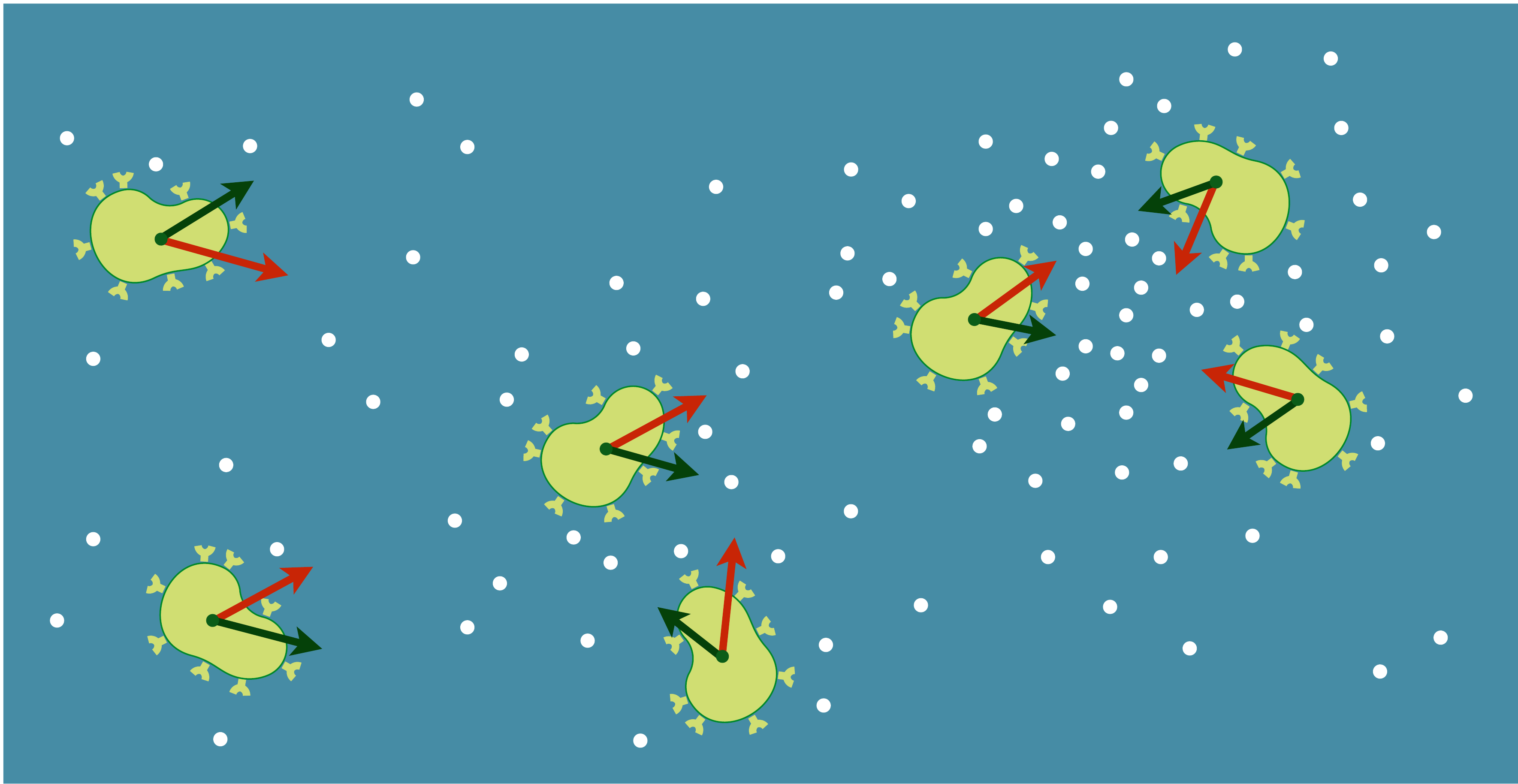}
\caption{Schematics of a system of chemotactic particles. The particles, shown as light green blobs, secret chemicals that are pictured as white dots. The red arrow at each point represents the chemotactic velocity corresponding to the $\bm{v}_{\rm KS}$ term in Eq.~\eqref{eq:cell_velocity}. The dark green arrows show velocities stemming from 
the  {polarity-induced} mechanism and correspond to the $\bm{v}_{\rm pi}$ term in the same equation.
}
\label{fig:schematic}
\end{figure}
%

We now focus on the dynamics of the cell polarity, $\mathbf{n}$. 
In general, the polarity of an isolated single cell undergoes a Brownian motion, and is randomized over the time scale of $D_r^{-1}$, where $D_r$ is an effective reorientation rate (akin to the rotational diffusion coefficient), which in our description represents the dominant mechanism for reorientation of the cell. This can be due to shape changes, cell polarity and cytoskeleton re-organization, solid rotation, etc.
In addition to this random process, it is also expected that a cell with many chemosensory units undergoes a 
 {polarity} change in response to an external chemical gradient. 
This can be achieved through shape changes, alignment via reorientation, or redistribution of surface receptors \cite{roussos2011,iglesias2008}.  
%
%
%
This form of response in polarity can be described by an effective angular velocity 
\begin{align} \label{eq:align}
\bm{\omega}=\chi \mathbf{n} \times \nabla \Phi, 
\end{align}
which is characterized by the 
 {polarity} coupling $\chi$, as demonstrated by the analysis of the representative toy model presented and discussed in  Appendix~\ref{app_toymodel}.

Note that, from Eq.~\eqref{eq:align}, the 
 {polarity} alignment happens over a time scale $\sim (\chi \nabla|\Phi|)^{-1}$. 
If this time scale is shorter than the reorientation time scale $D_r^{-1}$, a net bias in the average polarity of the cell along the direction of the gradient emerges beyond the polarity randomization time scale, which reads
(see Appendix~\ref{app:moment} for a detailed derivation):
\begin{equation}    \label{eq:n_handwaving}
\langle \mathbf{n} \rangle\approx \frac{\chi}{3 D_r} \nabla \Phi.
\end{equation}
Accordingly, by averaging the equations of motion for the position and for the polarity over time scales longer than the reorientation time and inserting Eq.~(\ref{eq:n_handwaving}) in the polar term in Eq.~(\ref{eq:v-nu1-nu2-1}), we obtain the following expression for the cell velocity
\begin{align} \label{eq:cell_velocity} 
    \frac{\mathrm{d}\mathbf{r}}{\mathrm{d}t}  = 
    \nu_1 \nabla \Phi +
    \frac{\nu_2 \chi}{6 D_r} \nabla (\nabla \Phi)^2
    + \bm{\xi}.
\end{align}
We observe that the two terms in Eq.~\eqref{eq:cell_velocity} depend differently on the properties of the chemical gradient vector field, as schematically represented in Fig.~\ref{fig:schematic}. Note that in this subsection we have only provided a heuristic derivation of these terms; a more systematic  {analysis is} presented in Appendix~\ref{app:moment}. 



\section{Stochastic model for generalized chemotaxis}	\label{sec:stoch}

In this section, we derive a stochastic field description for the dynamics of a system of chemotactic particles in $d$ spatial dimensions with both the KS and the
polarity-induced drift terms. 
In order to incorporate fluctuations into the description, 
we start from Eq.~\eqref{eq:cell_velocity} and
implement the Dean--Kawasaki (DK) approach to obtain a Langevin equation for the {\em instantaneous} particle density (denoted by $\hat{C}$) of the self-chemotactic system.  
We then phenomenologically extend this equation to include 
the possibility of particles switching between active (i.e,  responsive to the chemical fields) and inactive states.
Finally, we expand the extended DK equation around a uniform particle density $C_0$, which represents a homogeneous state, and obtain a Langevin equation for the fluctuations of the particle density. 
We then discuss the symmetry properties of the resulting dynamics, and the different states of the system that it describes.

%
%
\subsection{Stochastic conserved evolution equation for generalized chemotaxis using the Dean--Kawasaki approach}
\label{sec:DK}

To obtain the dynamics of an assembly of chemotactic cells starting from their microscopic dynamics as given by Eq.~\eqref{eq:cell_velocity}, one can use the DK approach~\cite{dean1996,kawasaki1994} 
and derive the exact stochastic dynamics of the instantaneous density field defined as
\begin{align}   \label{eq:C_exactdef}
\hat{C} (\mathbf{x},t) = \sum_a \delta( \mathbf{x}-\mathbf{r}_a),
\end{align}
where $\mathbf{r}_a$ indicates the position of the $a$-th particle of the assembly. 
The exact Langevin equation for $\hat{C}$ (or a smoothed version of it obtained upon coarse-graining~\cite{archer2004})
is then given by the continuity equation~\cite{dean1996}:
\begin{align} \label{eq:continuity_DK}
\partial_t \hat{C} (\mathbf{x},t) 
+ \nabla\cdot \mathbf{J}_{\rm DK} (\mathbf{x},t) = 0.
\end{align}
The instantaneous particle current $\mathbf{J}_{\rm DK}$, which encompasses the diffusion of the particles as well as the chemotactic interactions among them, is given by
\begin{align}       \label{eq:J_DKdef}
\mathbf{J}_{\rm DK} = -D\nabla \hat{C} (\mathbf{x},t) 
&+ \hat{C}(\mathbf{x},t) 
\left\lbrace \nu_1 \nabla \hat{\Phi} + \frac{\nu_2 \chi}{6 D_r} \nabla(\nabla \hat{\Phi} )^2 \right\rbrace \nonumber\\
& + \sqrt{ \hat{C} (\mathbf{x},t)} \, \bm{\xi}(\mathbf{x},t).
\end{align}
Note that the microscopic noise $\bm{\xi}(t)$ in Eq.~\eqref{eq:cell_velocity} has led to the Gaussian noise field $\bm{\xi}(\mathbf{x},t)$  in the particle current (we have 
kept the same notation for simplicity), with 
$\left\langle \bm{\xi}(\mathbf{x},t) \right\rangle = \mathbf{0}$
and
\begin{align}   \label{eq:xinoisecorr}
\left\langle \xi_i(\mathbf{x},t) \xi_j(\mathbf{x}',t') \right\rangle = 2 D\delta_{ij} \delta^d(\mathbf{x}-\mathbf{x}') \delta(t-t'),
\end{align}
where $D$ is the particle diffusion coefficient.

We now consider the dynamics of the chemical field~$\Phi$. 
In a self-chemotactic system---which is the focus of our work and is defined as a system of particles that produce and/or consume chemicals which they chemotactically respond to---the instantaneous chemical field $\hat{\Phi}(\mathbf{x},t)$ is continuously created by the diffusing chemical molecules that the particles release. The concentration of these chemicals is thus governed by a diffusion equation where the instantaneous particle density $\hat{C} (\mathbf{x},t)$ is the time-dependent source. 
Due to their size difference, the diffusion constant of the chemical molecules is often $10^2 - 10^3$ times larger than that of the particles secreting them~\cite{hofer1995, luca2003}. 
We hence assume that the chemical field $\hat{\Phi}$ instantaneously reaches the steady-state profile corresponding to a given $\hat{C}(\mathbf{x},t)$  
which is therefore governed by
\begin{align} 	\label{eq:Phi_exactPoisson}
\left(-\nabla^2 + \kappa^2\right) \hat{\Phi} (\mathbf{x},t) 
= 
\hat{C} (\mathbf{x},t) \, , 
\end{align}
where $\kappa^{-1}$ sets an effective screening length. 
This length scale is determined by the competition between the diffusion of the chemicals and a decay rate that is either due to the degradation of the chemical signals, in which case the length scale is typically much larger than the system size, or out-fluxes at the system boundary, in which case it will be comparable to the system size. 
%
It is worth mentioning that in addition to the effect on sensing and motion, the polarity of the particles can influence the production of the chemicals \cite{saha2014}. 
Such effect will, for example, lead to a depolarization effect akin to that observed in dielectric materials \cite{golestanian2019}. 
To a good approximation, the consequence of the anisotropic chemical release can be taken into account via a renormalization of the relevant parameters. Therefore, this feature is ignored in the current study for simplicity.


Finally, we note that when the chemical field $\Phi$ is governed by Eq.~\eqref{eq:Phi_exactPoisson}, the  KS contribution to the particle current~\eqref{eq:J_DKdef} 
can be derived from a free energy functional~\cite{chavanis2010,dean1996}, i.e.,
\begin{align}   
    \mathbf{J}_{\rm DK} = 
    - \hat{C} \nabla \frac{\delta \mathcal{F}_{\rm KS}}{\delta \hat{C} }  
    + \frac{\nu_2 \chi}{6 D_r} \hat{C} \nabla (\nabla \hat{\Phi} )^2 
    + \sqrt{ \hat{C} } \,\bm \xi (\mathbf{x},t)  ,
\end{align}
with the functional
\begin{align}
    \mathcal{F}_{\rm KS} [\hat{C}] = 
    D \int {\rm d}^d\mathbf{x} \, \hat{C} \log \hat{C} 
    - \frac{\nu_1}{2} \int {\rm d}^d\mathbf{x} \, \hat{C} \hat{\Phi} , \label{eq:KSfunctional}
\end{align}
 whereas the contribution from the polarity-induced mechanism (the $\frac{\nu_2 \chi}{6D_r}$ term in Eq.~\eqref{eq:J_DKdef}) \textit{cannot} be derived from such a functional (see Appendix~\ref{app_detailedbalance}).
This means that the KS part of the current is an equilibrium-like contribution, and, together with the corresponding noise, it satisfies the condition of detailed-balance.
On the other hand, the contribution from the polarity-induced mechanism for chemotaxis introduces a genuine nonequilibrium interaction between the particles .

%
%
\subsection{Extension to the nonconserved case with linear growth term}
\label{sec:nonconserved}

In the DK approach, the system under consideration has a constant number of particles and therefore the corresponding stochastic field equation, Eq.~\eqref{eq:continuity_DK}, takes the form of a conservation law. This description, however, does not take into account the possibility for changes in the chemical activity of the particles.
It has been observed that some chemotactic particles, such as fibroblasts \cite{kay2008}, can switch between active and inactive states.
We include this possibility within our model and phenomenologically extend the DK equation by introducing terms that do not conserve the number of particles.

The microscopic processes that activate and inactivate the chemotactic response of the particles can be represented via the stochastic process
\begin{align}       \label{eq:activedormantmicro}
   \mathrm{ACTIVE} \xrightleftharpoons[\,\, \lambda' \,\,]{\,\, \lambda \,\,} \mathrm{INACTIVE},
\end{align}
where $\lambda$ and $\lambda'$ represent the corresponding rates. Here, we have assumed that the inactive particles are abundant and their concentration remains effectively constant. This is a reasonable assumption for the case of fibroblasts, for example, which are activated only in response to local inflammation or cancerous activity \cite{yeo2018}. 

Using a system size expansion approach~\cite{vankampen1992,gillespie2000}, we obtain the stochastic field equation corresponding to these processes. Combining this with Eq.~\eqref{eq:continuity_DK}, we find the required phenomenological extension as 
\footnote{It should be noted this description is phenomenological and valid at the level of a density field which is coarse-grained over a spatial region but keeps track of the fluctuations (often denoted by $\bar{C}$). For detailed discussion see  Refs.~\cite{archer2004,chavanis2010}. For the sake of simplicity,  we have kept the same notation for both instantaneous density of the DK approach and this coarse-grained density field. 
}
\begin{equation}   \label{eq:lingrowth+DK}
\begin{split}
   \partial_t \hat{C} (\mathbf{x},t) + \nabla\cdot {\mathbf{J}} = &-\lambda \left[ \hat{C} (\mathbf{x},t) - C_0 \right] \\
  & + \sqrt{ \hat{C} (\mathbf{x},t) + C_0} \, \eta(\mathbf{x},t) \, , 
\end{split}
\end{equation}
where we have defined $C_0 = \lambda'/\lambda$ and the white noise $\eta$ is characterized by
\begin{align}   \label{eq:etanoisecorr}
    \left\langle \eta(\mathbf{x},t) \eta(\mathbf{x}',t')\right\rangle = 2 \lambda \, \delta^d\left(\mathbf{x}-\mathbf{x}'\right)\delta\left(t-t'\right),
\end{align}
and assumed to be uncorrelated with $\bm{\xi}(\mathbf{x},t)$.
Note that for $\lambda>0$, the growth term in Eq.~\eqref{eq:lingrowth+DK} tends to drive the system towards a homogeneous configuration with uniform density $C_0$, which can be considered as the homeostatic state of the system.

%
\subsection{Stochastic field equation for the fluctuations of particle density around a homogeneous state}
\label{sec:langevin}

Following the previous section, we now consider systems where the particle density can be written as fluctuations around the uniform value $C_0$, and aim to obtain the equations that govern the dynamics of the density fluctuations. In particular, we assume
\begin{align}	\label{eq:Chat_expansion}
\hat{C} (\mathbf{x},t) = C_0 + \rho(\mathbf{x},t) \, , \quad \mathrm{with} \quad |\rho(\mathbf{x},t)| \ll C_0
\end{align}
where $\rho$ represents the density fluctuations around $C_0$. Note that $\rho$ does not need to stay positive and the assumption of small fluctuations ensures the positivity of the total particle density $C$ at all times.
In order to study the dynamics of $\rho$, both Eq.~\eqref{eq:Phi_exactPoisson} for the chemical concentration field and the extended DK description of the particle density given by Eq.~\eqref{eq:lingrowth+DK} should be expanded using Eq.~\eqref{eq:Chat_expansion}.

Similarly, the chemical field is expanded as 
$\hat{\Phi}(\mathbf{x},t) = \Phi_0 + \phi(\mathbf{x},t)$ where $\Phi_0$ 
is the base value of the chemicals maintained by the uniform part of the particle density, and $\phi(\mathbf{x},t)$ represents the chemical fluctuations caused by the fluctuations of the particle density. Substituting this in Eq.~\eqref{eq:Phi_exactPoisson} gives
\begin{subequations}    \label{eq:screenedPoissonSeperation}
\begin{align}
(-\nabla^2 + \kappa^2)\, \Phi_0 &= C_0 \, ,\\
(-\nabla^2 + \kappa^2)\,\,\, \phi\, &= \rho(\mathbf{x},t) \, ,
\end{align}
\end{subequations}
where the first equation has a uniform solution 
$\Phi_0 = \kappa^{-2} C_0$.
Any gradients in the chemical concentration can thus only be due to the chemical fluctuation field~$\phi$. 
%
%
Taking the limit where the characteristic length scale for variations in the system is much smaller $\kappa^{-1}$, which corresponds to situations where the chemical signals do not decay considerably within the system size or observation scale, gives the Poisson equation 
\begin{align}	\label{eq:poisson}
-\nabla^2\phi(\mathbf{x},t) =  \rho(\mathbf{x},t).
\end{align}

To expand the extended DK equation \eqref{eq:lingrowth+DK}, we first rewrite the DK current \eqref{eq:J_DKdef} by using Eq.~\eqref{eq:Chat_expansion}, which yields
\begin{equation}    \label{eq:JDK_fullexpansion}
    \begin{split}
            \mathbf{J}_{\rm DK} = &-\!D\nabla\rho + C_0 \nu_1 \nabla\phi + C_0 \frac{\nu_2 \chi}{6 D_r} \nabla\left(\nabla\phi\right)^2  + \nu_1 \rho\nabla\phi \\
            &+ \frac{\nu_2 \chi}{6 D_r} \rho\nabla\left(\nabla\phi\right)^2 + \sqrt{C_0}\, \sqrt{1 + \frac{\rho}{C_0}} \, \, \bm{\xi}\left(\mathbf{x},t\right).
    \end{split}
\end{equation}
Note that as a result of the Poisson equation \eqref{eq:poisson}, both $\nabla\left(\nabla\phi\right)^2$ and $\rho\nabla\phi$ have similar scaling as $\rho^2$ (in the dynamical equation), while $\rho\nabla\left(\nabla\phi\right)^2$ scales as $\rho^3$ and, henceforth, will be discarded as a higher order term for small density fluctuations (for a detailed discussion of this approximation see Sec.~\ref{sec:symmetry}).
Substituting the resulting expression for $\mathbf{J}_{\rm DK}$ into Eq.~\eqref{eq:lingrowth+DK} and expanding the remaining terms gives the following extension of the stochastic KS model 
\begin{equation}
    \label{eq:langevin}
    \begin{split}
    ( \partial_t - D \nabla^2 + \sigma) \, \rho (\mathbf{x},t) = &- \mu_1{\nabla} \cdot (\rho\nabla\phi)\\
    &- \mu_2\nabla^2(\nabla\phi)^2 + \zeta(\mathbf{x},t)\,,
    \end{split}
\end{equation}
where in terms of the microscopic parameters we have
\begin{equation}
\sigma=\lambda-C_0 \nu_1,
\quad
\mu_1 = \nu_1,
\quad
\mu_2 = C_0 \frac{\nu_2 \chi}{6 D_r}.
\end{equation}
Moreover, the noise field $\zeta(\mathbf{x},t)$ is obtained by keeping only the additive parts of the original noise fields $\bm{\xi}$ and $\eta$ when expanding the density around $C_0$ (see the remarks below), and reads as
\begin{align}   \label{eq:zetanoisedef}
\zeta(\mathbf{x},t) = -\sqrt{C_0} \,\, \nabla\cdot\bm{\xi}(\mathbf{x},t) + \sqrt{2 C_0} \,\, \eta(\mathbf{x},t). 
\end{align} 
The corresponding correlations are calculated as 
\begin{equation}	\label{eq:zetanoisecorr}
\left\langle\zeta(\mathbf{x},t)\zeta(\mathbf{x}',t')\right\rangle = 2(\mathcal{D}_0-\mathcal{D}_2\nabla^2)~\delta^d(\mathbf{x}-\mathbf{x}')\delta(t-t')\,,
\end{equation}
where $\mathcal{D}_0 = 2 C_0 \lambda$ and $\mathcal{D}_2 = C_0 D$ in terms of the microscopic parameters. 

The stochastic field equation~\eqref{eq:langevin} is the main result of this section, and in the rest of the paper we will analyze its mathematical structure and scaling behavior. 
A few pertinent remarks regarding this equation shall be mentioned below. 

First, for $\mathcal{D}_0 \neq 0$ the noise does not conserve 
the number of particles and this applies to the cases where number fluctuations are allowed 
(see discussion in Sec.~\ref{sec:nonconserved}). 
When $\mathcal{D}_0 = 0$, on the other hand, the resulting conserved noise induces fluctuations only in the particle current $\mathbf{J}_{\rm DK}$ and therefore $\rho$ is locally conserved, as it happens in systems with strictly fixed number of particles, e.g., active colloids. 

Second, we note that in addition to the additive noise~$\zeta$ defined in Eq.~\eqref{eq:zetanoisedef}, the expansion of Eq.~\eqref{eq:lingrowth+DK} contains also multiplicative noise terms with correlations proportional to (positive) powers of $\rho$.
In Sec.~\ref{sec:RG}, we show that these multiplicative terms are \textit{irrelevant} in the RG sense and can be neglected when analyzing the critical behavior of the system. 
It should be emphasized that the assumption of $C_0\neq0$ is crucial here as it allows the expansion around the additive noise. For $C_0 \to 0$, the additive part of the noise vanishes and, consequently, the multiplicative terms cannot be discarded anymore. The investigation of this case and the possible transition to an absorbing state of the system is left for future work.

%
%
%
%
\subsection{Galilean symmetry}
\label{sec_galilean}

Before studying the critical regime of the stochastic field equation \eqref{eq:langevin} derived in the previous sections, we first discuss here its relevant emerging symmetry. Consider the Galilean transformation defined by
\begin{subequations} \label{eq:galileantransform}
\begin{align}
\phi'(\mathbf{x},t) &= \phi\big(\mathbf{x}+(\mu_1-2\mu_2) t\mathbf{w},t\big)-\mathbf{w}\cdot\mathbf{x},\\
\rho'(\mathbf{x},t) &= \rho \big(\mathbf{x}+(\mu_1-2\mu_2) t\mathbf{w},t\big) \, , 
\end{align}
\end{subequations}
where $\mathbf{w}$ is an arbitrary $d$-dimensional vector. Under this transformation of the fields, and noting the Poisson equation~\eqref{eq:poisson}, the stochastic field equation~\eqref{eq:langevin} remains invariant. This symmetry plays a crucial role in our following analysis, since it constrains the nonlinear couplings that can be generated by the RG flow and yields an exact identity between the critical exponents, as we discuss in the following. 
Although this symmetry is not present at the microscopic level, we emphasize that it emerges when the diffusion of the chemical signals is considerably faster than that of the particles, and when the  {screening length set by the decay rate of the chemicals is larger than the characteristic length scales in the system, leading to Eq.~\eqref{eq:poisson}}. 
Note that this symmetry remains valid since the noise is delta-correlated in time~\cite{medina1989}.
%
%

%
%
%
%
\subsection{Dispersion, collapse, and the critical state}
\label{sec:criticalpoint}

\begin{figure}[t]
	\centering
		\includegraphics[width=0.48\textwidth]{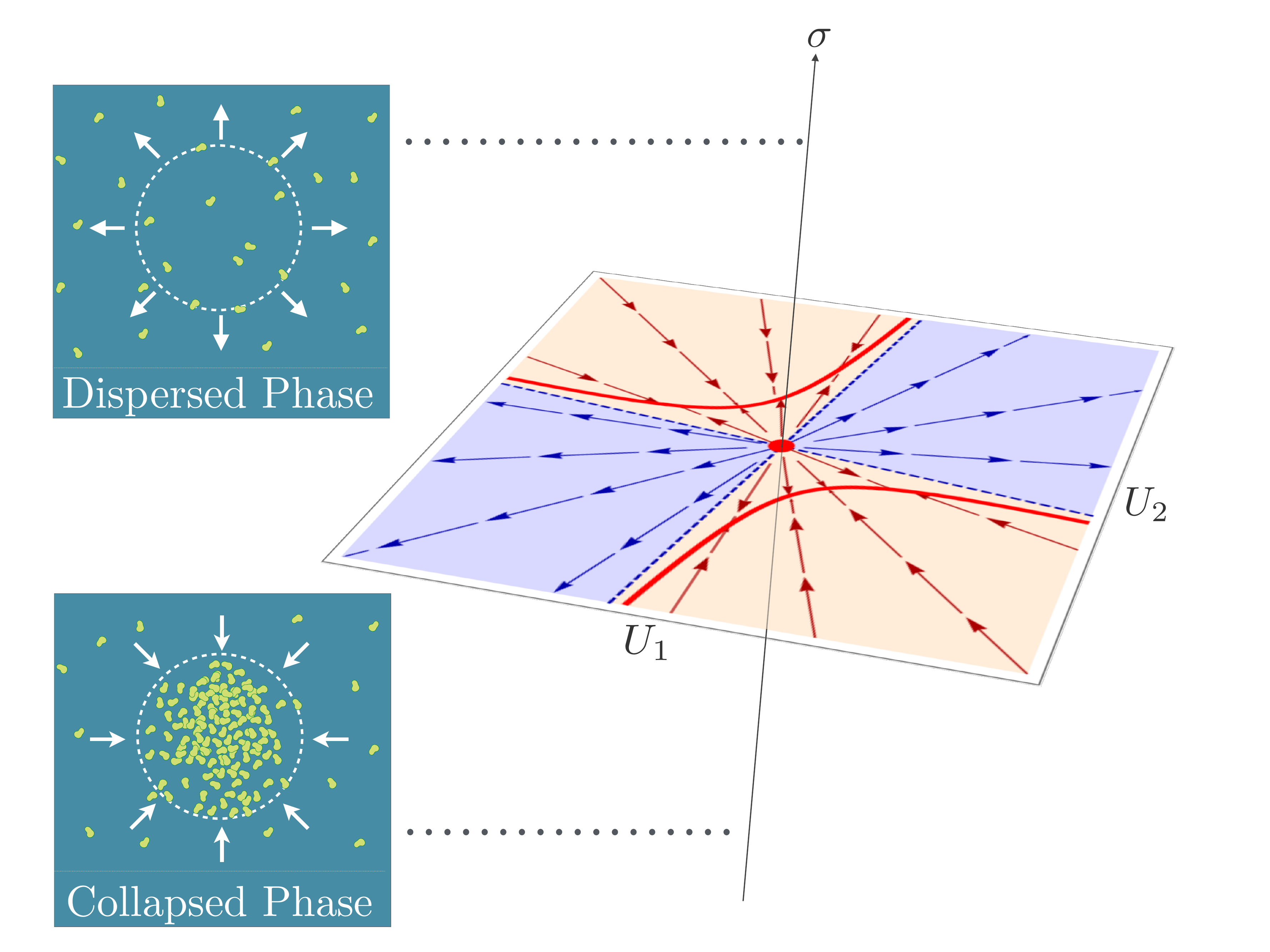}
	\caption{Phase diagram of the chemotactic model described in Eq.~\eqref{eq:langevin}. The control parameter $\sigma$ delimits two phases: a dispersed phase for which the density fluctuations around $C_0$ are exponentially suppressed when $\sigma>0$, and a collapsed phase when $\sigma<0$, in which particles are attracted to regions of high concentration. 
When $\sigma=0$, the system is critical and its long-time and large-scale behavior is described using the renormalization group framework developed here.}
	\label{fig:phaseTransition}
\end{figure}

The competition between the KS and the 
polarity-induced chemotactic interactions, and the linear growth terms, determines the properties of the fluctuations at long times. This competition is reflected in Eq.~\eqref{eq:langevin} through the sign of the parameter $\sigma$, which within our microscopic description is given by $\sigma=\lambda-C_0 \nu_1$: when $\sigma>0$, or equivalently $\nu_1< C_0^{-1} \lambda$, the density fluctuations $\rho$ tend to decay exponentially with time, whereas for $\sigma<0$, or $\nu_1>C_0^{-1} \lambda$, the fluctuations tend to grow. 
Microscopically, these relationships show that when chemotaxis is relatively weak, modulations of the density profile around the homogeneous state are automatically smoothed out and the system returns to the uniform profile, whereas for large chemotactic interactions the perturbations are amplified, resulting in a collapsed state \cite{jager1992}, see the phase diagram on Fig.~\ref{fig:phaseTransition}. 

When $\sigma$ vanishes, on the other hand, the density fluctuations become long-lived, and the correlation length of Eq.~\eqref{eq:langevin}, which is set by $\sqrt{D/\sigma}$ in the Gaussian approximation, diverges. Therefore, the \textit{critical state} of this system is reached by fine-tuning $\sigma$ to zero which, in the microscopic description, can only be done for $\nu_1>0$ when $\lambda>0$. 
Note that the connection between the coupling constants in the coarse-grained theory with their microscopic values is, however, nontrivial and should be established via a renormalization procedure. One can, in general, consider Eq.~\eqref{eq:langevin} to hold at the macroscopic level with the parameters having arbitrary values independent of each other. 

%
%

\section{Renormalization group analysis}	\label{sec:RG}

In this section we investigate the critical state of the system ($\sigma=0$) by first analyzing the scaling properties of the particle density and then employing RG techniques to study the RG flow of the coupling constants due to the coarse-graining and rescaling of the theory. This section, together with Appendix~\ref{app_rgprocedure}, contains the technical details of the RG analysis for the interested reader. It can be skipped if the reader wishes to focus on the important results, which are presented in Sec. \ref{sec:results}.

The critical state with $\sigma = 0$ lies in between the stable dynamics with $\sigma>0$, where $\rho$ decays to zero exponentially in time, and the  unstable region where $\rho$ grows exponentially. 
In both cases, a mean-field treatment is sufficient to understand the macroscopic physics beyond the correlation length $\sim\sqrt{D/\sigma}$. 
On the other hand, in the critical region the correlations are long-ranged and the fluctuations in the particle density are long-lived, hence the fluctuations in the particle density interacting via the chemotactic couplings determine the macroscopic properties. This will also hold when the decay length as set by $\sigma$ is considerably larger than the system size. RG techniques can then be utilized to gain valuable information about the macroscopic properties and the corresponding phase transitions for different values of the chemotactic couplings.  


\subsection{Scaling analysis and upper critical dimensions}
\label{sec_scaling}


At the critical point, the correlation lengths of the solutions of the nonlinear Langevin dynamics Eq.~\eqref{eq:langevin} diverge and therefore the theory becomes scale invariant~\cite{tauber2014}. We consider the scaling behavior of the critical solutions under a change of the spatial and temporal scales given by
\begin{align}
    \mathbf{x}' = \mathbf{x}/b \, , \quad \text{and} \quad  t' =t/b^z \, ,
    \label{eq_scaling}
\end{align}
where $b>1$ is the scaling factor, and a corresponding scaling of the particle density fluctuations and chemical fluctuations as 
\begin{align}
\rho' = \rho/b^\chi\, , \quad \text{and} \quad \phi' = \phi/b^\psi \, ,
\end{align}
respectively.  We have introduced three scaling exponents: $\chi$, often known as the ``roughness'' exponent in the context of surface growth dynamics\footnote{Note that one could alternatively introduce Fisher's anomalous exponent $\eta$, with $\chi = -(d+\eta)/2$ in the conserved case and $\chi = (2 - d - \eta)/2$ in the nonconserved case.}, 
the dynamic exponent $z$, and the chemical field exponent $\psi$. 
These scaling exponents are not fully independent due to the relationships between the physical variables of the system: in our case, the chemical exponent $\psi$ is related to the roughness exponent $\chi$ by the Poisson equation~\eqref{eq:poisson}, which yields 
\begin{align}
\psi=\chi+2. 
\end{align}
Furthermore, the Galilean symmetry, Eq.~\eqref{eq:galileantransform}, imposes another exponent identity: $\phi$ scales as 
$\mathbf{w}\cdot \mathbf{x}$, which yields $\psi=2-z$ if the the nonlinearities $\mu_{1,2}$ are taken to be dimensionless, which is consistent with $z+\chi=0$, since we have $\psi=\chi+2$ (see Sec.~\ref{sec:expo} for more discussion). 

In the absence of the nonlinear terms in Eq.~\eqref{eq:langevin}, the values of the exponents introduced above can be obtained by requiring the invariance of the equation under the change of spatial and temporal scales~\eqref{eq_scaling}. Depending on whether the noise is conserved ($\mathcal{D}_0=0$) or nonconserved ($\mathcal{D}_0\neq0$), we obtain the following Gaussian dimensions for the density (fluctuations) field: 
\begin{subequations}    \label{eq_fieldDimensions}
\begin{align}
    \chi_{0}^{\rm con} &= -\frac{d}{2} \, , \quad \text{for conserved noise} \, , \\
    \chi_{0}^{\rm non} &= \frac{2-d}{2} \, , \quad \text{for nonconserved noise} \, ,
\end{align}     
and the dynamic exponent takes the value 
\begin{align}
z_{0}=2,
\end{align}
\end{subequations}
in both cases. 
Based on these engineering dimensions, a dimensional analysis reveals that with a conserved noise, the nonlinear terms $\mu_{1,2}$ scale as 
$\propto b^{2-d/2}$ at the Gaussian fixed point and hence grow upon successive applications of the rescaling procedure if $d<d^{\rm con}_c=4$. For the nonconserved noise, on the other hand, the nonlinearities scale as $\propto b^{3-d/2}$ at the Gaussian fixed point and grow in $d<d^{\rm non}_c=6$ spatial dimensions. Accordingly, below the critical dimension $d_c$ the nonlinearities $\mu_{1,2}$ are \textit{relevant} in determining the scaling behavior of the system and, therefore, need to be examined via the RG analysis. 
(A systematic discussion based on power counting is given in Appendix~\ref{app_powerCounting}.)

As the final remark, we turn to the scaling properties of the noise terms and note that in the presence of a nonconserved noise ($\mathcal{D}_0\neq0$), the conserved noise has a scaling dimension equal to $-2$, and is therefore irrelevant and can be discarded from the analysis of the critical state. Furthermore, because of the scaling of $\rho$ determined by Eq.~\eqref{eq_fieldDimensions}, the multiplicative noise terms with correlations proportional to $\rho^n$ have an engineering dimension given by $-n d/2$ in the conserved case, and $n(1-d/2)$ in the nonconserved case. Such terms are therefore irrelevant at the upper critical dimension in both cases and this justifies discarding them from Eq.~\eqref{eq:langevin} in the analysis of the critical regime.

%
%
%
%
\subsection{Renormalization group flow equations}
\label{sec_RG}

Below the upper critical dimension $d_c$, the nonlinearities in Eq.~\eqref{eq:langevin} are relevant, and we implement a perturbative momentum-shell renormalization group procedure~\cite{forster1977,medina1989,tauber2014} to study the critical behavior of the chemotactic particles. This procedure is conveniently implemented in the Fourier space, where upon performing Fourier transformations according to 
\begin{align}   \label{eq:fourierdef}
\rho(\mathbf{x},t)=\int_{\hat{k}} e^{-i\omega t+i\mathbf{k}\cdot\mathbf{x}} \rho(\hat{k})
\end{align}
with 
$\hat{k}=(\mathbf{k},\omega)$ 
and 
$\int_{\hat{k}}\equiv\int \mathrm{d}\omega \, \mathrm{d}^d \mathbf{k}/(2\pi)^{d+1}$, 
and using Eq.~\eqref{eq:poisson} to represent $\phi$ in terms of $\rho$, Eq.~\eqref{eq:langevin} reads as
\begin{align}   \label{eq:fourierlangevin}
\rho(\hat{k})= G_0(\hat{k})\left[\zeta(\hat{k}) + \int_{\hat{q}} \Gamma_0(\mathbf{k},\mathbf{q}) \rho(\hat{k}-\hat{q}) \rho(\hat{q}) \right]. 
\end{align}
Here we have introduced the \textit{bare} propagator~$G_0$:
\begin{align}    \label{eq:barepropagator}
G_0(\hat k) &= \left(\sigma-i\omega+D\mathbf{k}^2\right)^{-1} =
\begin{tikzpicture}
[baseline={([yshift=-.5ex]current bounding box.center)},vertex/.style={anchor=base,
    circle,fill=black!25,minimum size=18pt,inner sep=2pt}]
\draw [line width=0.3mm, -{Latex[length=2.5mm,reversed]}] (-0.7,0) -- (0,0);
\draw [line width=0.3mm] (-0.2,0) -- (0.5,0);
\node at (-0.15,0.4) {$\hat{k}$};
\node at (0,-0.6) {$ $};
\end{tikzpicture} 
\, ,
\end{align}
and the bare (chemotactic) interaction vertex $\Gamma_0$ as:
\begin{align}        \label{eq:barevertex}
\Gamma_0(\mathbf{k},\mathbf{q}) &= \frac{\mu_1}{2} \left(\frac{\mathbf{k}\cdot\mathbf{q}}{\mathbf{q}^2} + \frac{\mathbf{k}\cdot(\mathbf{k}-\mathbf{q})}{(\mathbf{k}-\mathbf{q})^2}\right)-\mu_2 \frac{\mathbf{k}^2 \mathbf{q}\cdot(\mathbf{k}-\mathbf{q})}{\mathbf{q}^2(\mathbf{k}-\mathbf{q})^2} \nonumber  \\
&=\begin{tikzpicture}
[baseline={([yshift=-.5ex]current bounding box.center)},vertex/.style={anchor=base,
    circle,fill=black!25,minimum size=18pt,inner sep=2pt}]
\draw [line width=0.3mm] (-1,0) -- (0,0);
\draw [line width=0.3mm, -{Latex[length=2.5mm,reversed]}] (-1,0) -- (-0.4,0);
\draw [line width=0.3mm] (0,0) -- (45:1.0cm);
\node at (-0.6,.4) {$\mathbf{k}$};
\draw [line width=0.3mm, -{Latex[length=2.5mm,reversed]}] (0,0) -- (45:.75cm);
\node at (.8,1) {$\mathbf{q}$};
\draw [line width=0.3mm] (0,0) -- (-45:1.0cm);
\draw [line width=0.3mm, -{Latex[length=2.5mm,reversed]}] (0,0) -- (-45:.75cm);
\node at (.8,-1) {$\mathbf{k}-\mathbf{q}$};
\end{tikzpicture} 
\, .
\end{align}%
\charlie{%
Note that the vertex is symmetric with respect to the two in-coming legs:
$\Gamma_0(\mathbf{k},\mathbf{q})=\Gamma_0(\mathbf{k},\mathbf{k}-\mathbf{q})$.
}
In addition, we also define the bare dynamic correlation function as 
\begin{align}   \label{eq:barecorr}
\mathcal{N}_0 (\hat k) &= 2(\mathcal{D}_0+k^2\mathcal{D}_2) |G_0(\hat k)|^2 = 
\begin{tikzpicture}
[baseline={([yshift=-.5ex]current bounding box.center)},vertex/.style={anchor=base,
    circle,fill=black!25,minimum size=18pt,inner sep=2pt}]
\draw [line width=0.3mm] (-.5,0) -- (0,0);
\draw [line width=0.3mm, -{Latex[length=2.5mm,reversed]}] (-1,0) -- (-.4,0);
\draw [line width=0.3mm] (0,0) -- (1,0);
\draw [line width=0.3mm, {Latex[length=2.5mm,reversed]}-] (0.4,0) -- (1,0);
\node[circle,draw=black, fill=white, inner sep=0pt,minimum size=5pt] (b) at (0,0) {};
\node at (-0.6,0.4) {$\hat{k}$};
\node at (.6,0.4) {$-\hat{k}$};
\node at (-1,-0.6) {$ $};
\node at (.9,-0.6) {$ $};
\end{tikzpicture}
 \, . 
\end{align}
%
%
%
\begin{figure}[b]
	\centering
		\includegraphics[width=0.48\textwidth]{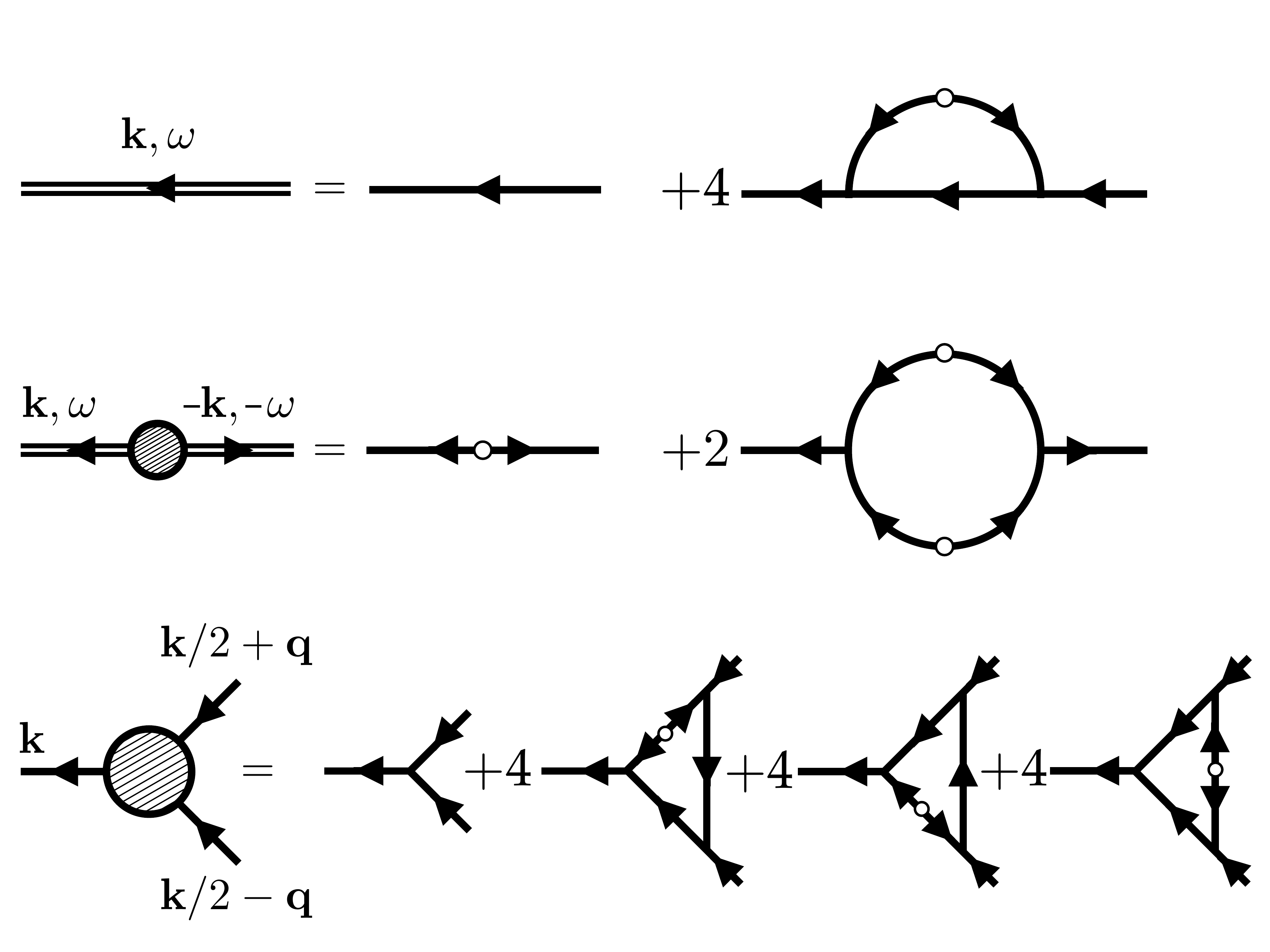}
	\caption{One-loop Feynman diagrams describing the renormalization of the propagator $G$ (top), the dynamic correlation function $\mathcal{N}$  (centre), and the vertex function $\Gamma$ (bottom) to the leading order.} 
	\label{fig:diagrams}
\end{figure}
%

\begin{figure*}[t]
	\centering
\hskip-1cm
	\begin{minipage}[c]{.34\linewidth}
		\centering
		\includegraphics[width=\linewidth]{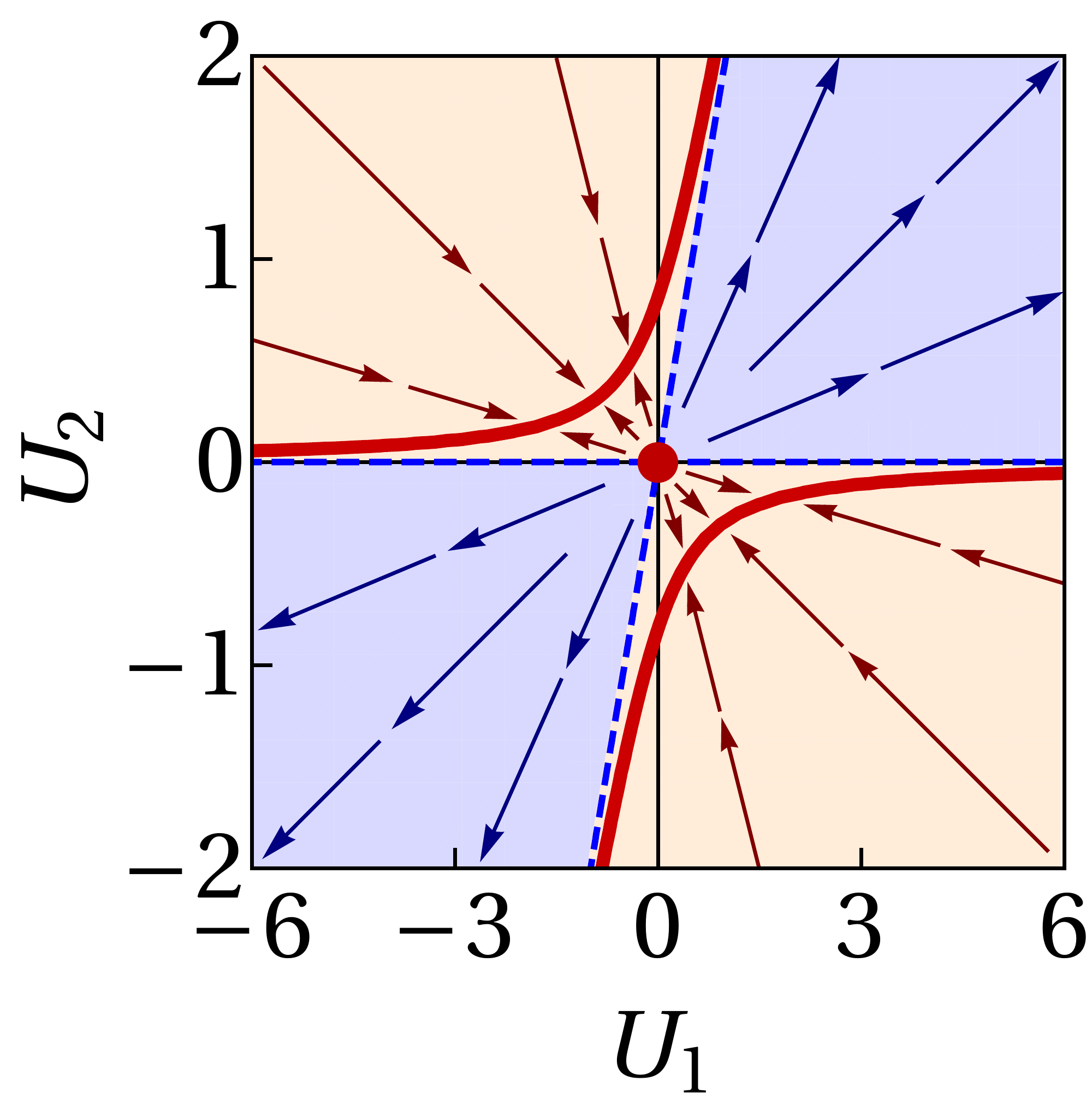}
	\end{minipage} 
	\hskip1cm
	\begin{minipage}[c]{0.34\linewidth}
		\centering
		\includegraphics[width=\linewidth]{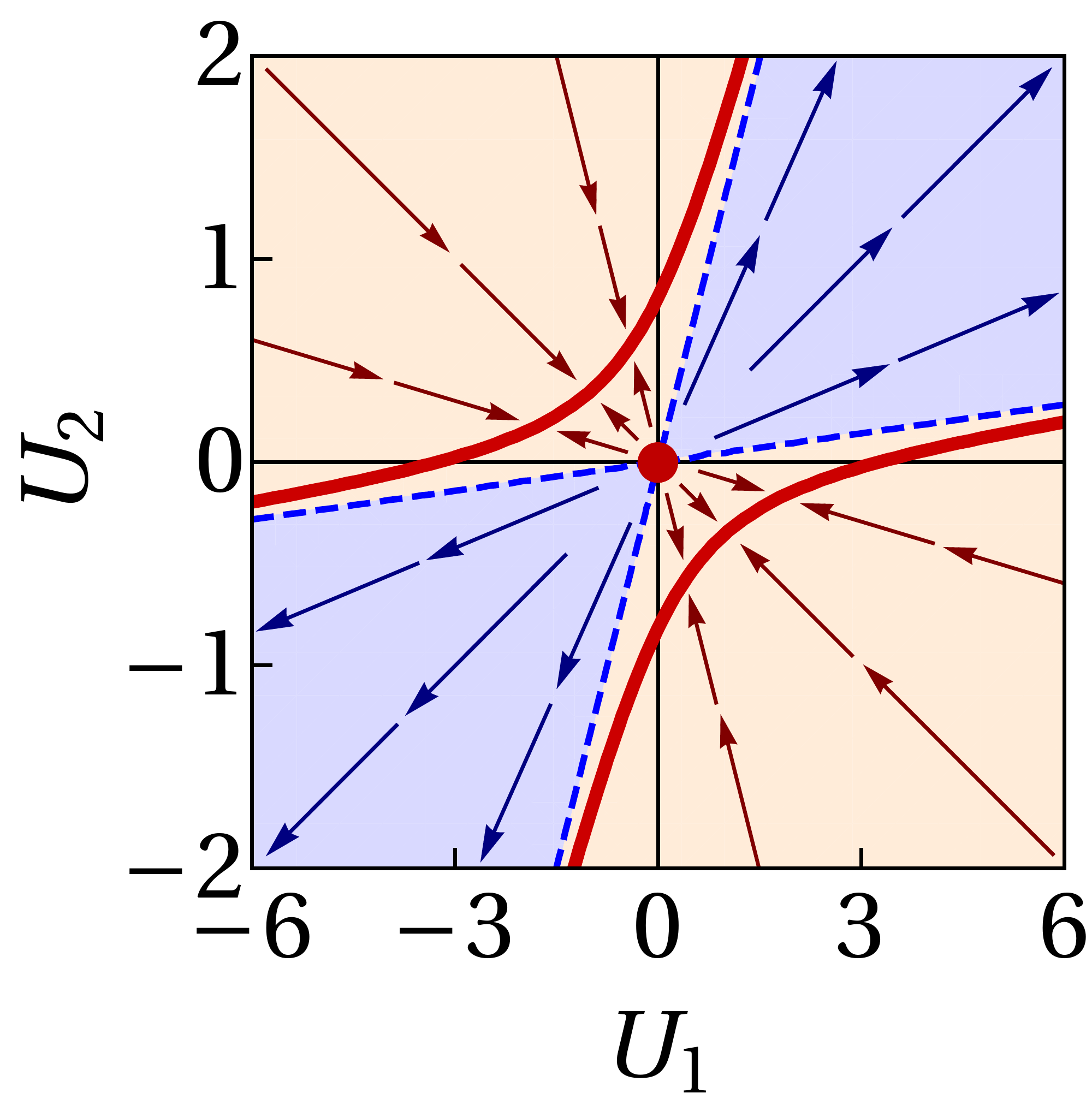}
		\end{minipage}
	\caption{RG flows in the $U_1$-$U_2$ plane in $d=2$ for conserved (left) and nonconserved (right) noise. 
	The arrows represent RG flows along rays passing through the origin, while the red solid lines are the hyperbolas of fixed points. Note that the $U_1$ axis has runaway flows for conserved noise but it has stable fixed-points for the nonconserved noise. The $U_2$ axis shows stable fixed points in both cases.
	} \label{fig:U1U2flows}
\end{figure*}

In the standard procedure \cite{medina1989,tauber2014}, the scale-invariant behavior of the system when the nonlinearities are relevant (i.e., below $d_c$) is captured by using perturbative RG techniques. A series expansion of Eq.~\eqref{eq:fourierlangevin} in terms of the couplings $\mu_{1,2}$ is constructed according to the Feynman diagrams in Fig.~\ref{fig:diagrams}. In the first step, the loop integrals that appear in the perturbation series are computed by integrating out the fluctuations whose wavevector $\mathbf{k}$ lies within the momentum shell $|\mathbf{k}|\in \left[\Lambda/b,\Lambda\right]$, where $\Lambda$ is the cutoff set by the microscopic length-scale of the particles. This step corresponds to coarse-graining of the short-distance fluctuations. 
In the second step, the variables are rescaled in a similar fashion to the mean-field case, so that the original cut-off $\Lambda$ is restored and the same Langevin equation~\eqref{eq:fourierlangevin} with renormalized (i.e., effective) coupling constants holds. 
Choosing an infinitesimal scaling parameter $b=e^\ell$ with $\ell \to 0$ casts the coarse-graining procedure into a differential form, which, in the case of the conserved noise, gives the following one-loop RG flow equations (see Appendix~\ref{app_rgprocedure} for details)
\begin{subequations}
\label{eq:cnoiseflows}
\begin{align}
\partial_{\ell} \sigma &= \left[ 2+d+2\chi \right] \sigma \label{eq:cnoiseflowsigma} ,\\ 
\partial_{\ell} \mu_{1,2} &=  \left[ z+\chi \right]\mu_{1,2}, \label{eq:cnoiseflowmu12}\\[1.5mm]
\partial_{\ell} D \,\, &= \Big[ z-2 - \big( a_{11} U_1^2 
     + a_{12} U_1 U_2 + a_{22} U_2^2\big)\Big] D,\label{eq:cnoiseflowD}
 \\[1.5mm]
\partial_\ell \mathcal{D}_2 &=  \left[-2-d+z-2\chi\right] \mathcal{D}_2, \label{eq:cnoiseflowD2}
\end{align}
\end{subequations}
with coefficients 
$a_{11} = 3/4 - 3/(2 d)$, 
$a_{12} = 2+3/d-6/(d+2)$, 
$a_{22} = 1-4/d$.
Here, we have defined the combined dimensionless chemotactic couplings 
$U_{1,2}^2 = \mu_{1,2}^2 \, \mathcal{D}_2 \, K_d \Lambda^{d-4}/D^3$ with $K_d=2/[(4\pi)^{d/2}\Gamma(d/2)]$.

We emphasize that the noise strength $\mathcal{D}_2$ and the chemotactic couplings $\mu_{1,2}$ are not renormalized in Eq.~\eqref{eq:cnoiseflows} and only the diffusion coefficient $D$ has a nontrivial RG flow.
For the noise term, one observes that the diagrams contributing to its renormalization include, at least, two bare vertices with external momenta $\mathbf k$ and $-\mathbf k$. In the limit $(k/q)\to 0$ taken for the shell integration, each of these bare vertices has the expansion
\begin{align}
    \Gamma_0(\mathbf k,\mathbf q) = -\mu_1 \frac{(\mathbf{k}\cdot\mathbf{q})^2}{q^4}+\left(\frac{\mu_1}{2}+\mu_2\right)\frac{k^2}{q^2} + \mathcal{O} \!\!\left(\frac{k}{q}\right)^{\!\!3} ,
    \label{eq_vertexExpansion}
\end{align}
which, upon multiplication, result in $(k/q)^4$ corrections to the noise term. 
However, since the noise term scales as $(k/q)^0$ in the nonconserved case and as $(k/q)^2$ in the conserved case, we conclude that the corrections are subleading and, hence, the noise is not renormalized in either case. 
For the chemotactic couplings $\mu_{1,2}$, on the other hand, the Galilean symmetry~\eqref{eq:galileantransform} directly imposes that $(\mu_1-2\mu_2)$ is not renormalized as it is the combination that appears in the symmetry transformation (see Appendix~\ref{app:ward} for a discussion of the associated Ward identity). 
In principle, the flows of $\mu_1$ and $\mu_2$ need not be vanishing separately and the fact that they both do not renormalize according to Eqs.~\eqref{eq:cnoiseflowmu12} may only be a one-loop result.

Rewriting Eq.~\eqref{eq:cnoiseflowD} in terms of the combined couplings $U_{1,2}$, we obtain
\color{black}
\begin{align}
\label{eq:cnoiseflowU1U2}
\frac{\partial_\ell U_{1,2}}{U_{1,2}} =  2-\frac{d}{2} 
+\frac{3}{2} \big(  a_{11}  U_1^2 + a_{12}  U_1 U_2 
+ a_{22} U_2^2 \big),
\end{align}
the solution of which traces rays with fixed $U_2/U_1$ in the $U_1$-$U_2$ plane, as shown in Fig.~\ref{fig:U1U2flows}. A similar analysis for the nonconserved noise, as outlined in Appendix~\ref{app_rgnon}, leads to RG flows analogous to Eq.~\eqref{eq:cnoiseflows} and, upon introducing the suitable dimensionless chemotactic couplings 
$U_{1,2}^2 = \mu_{1,2}^2 \, \mathcal{D}_0 \, K_d \Lambda^{d-6}  / D^3$, they imply the flows
\begin{equation}
\label{eq:nnoiseflowU1U2}
\frac{\partial_\ell U_{1,2}}{U_{1,2}} =  3-\frac{d}{2} 
+\frac{3}{2} \big(  b_{11}  U_1^2 + b_{12}  U_1 U_2 
+ b_{22} U_2^2 \big),
\end{equation}
with coefficients $b_{11} = 3/4 - 1/d - 3/[d(d+2)]$, $b_{12} = 2 + 6/d - 9/(d+2)$, and $b_{22} = 1-6/d$, 
which has the same structure as Eq.~\eqref{eq:cnoiseflowU1U2}.
%

\begin{figure*}[t]
   \centering
  \includegraphics[width=0.99\textwidth,keepaspectratio]{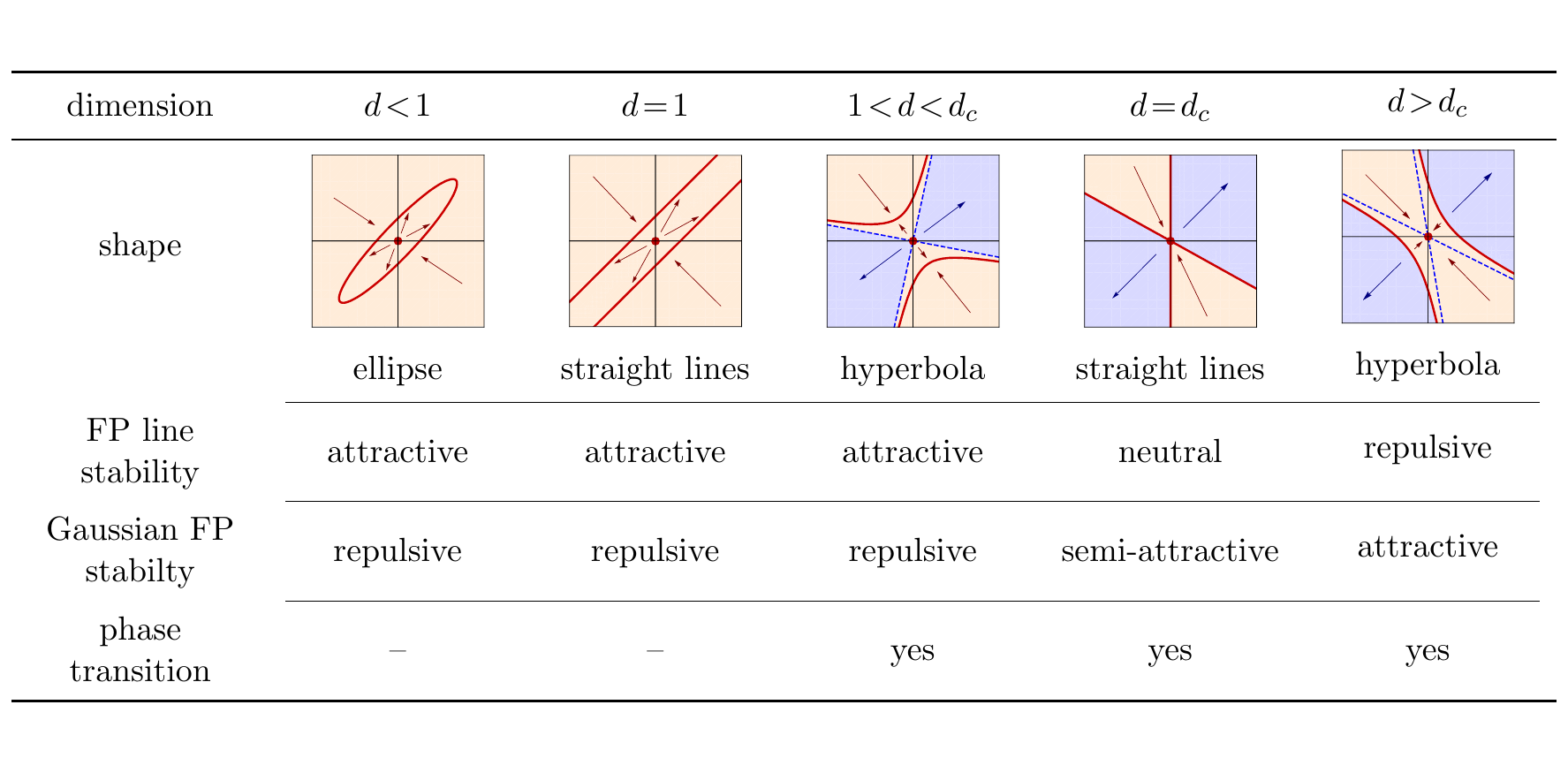}
    \caption{The structure of the RG fixed points (FP) in various dimensions (for $d=2$, see Fig.~\ref{fig:U1U2flows}). Note that in $d\leq 1$, the basin of attraction for the red curves is the whole plane while for $1<d<d_c$ the blue shaded regions show runaway flows. For $d>d_c$, the lines of nontrivial RG fixed points become repulsive (see Appendix~\ref{app_rgflowsdim} for the linear stability analysis).
  }
  \label{fig:rgflowvardim}
\end{figure*}
%
%

%
%
%
%
\section{Results} \label{sec:results}
\
This section addresses the outcome of the RG analysis of the Langevin equation~\eqref{eq:langevin} by, first, describing the RG flow diagrams that are obtained within our one-loop computation and, then, discussing the exact exponents that characterize the scaling laws of the critical system. {
\color{black}The role of the Galilean symmetry in obtaining the scaling exponents is reviewed, with some final remarks regarding possible symmetry-breaking terms in systems with self-propulsion and nematic anisotropy.}

%
%
\subsection{Flow diagrams}
\label{sec:phase}
%

The RG flows at one-loop order for the effective couplings $U_{1,2}$, given by Eqs.~\eqref{eq:cnoiseflowU1U2} and \eqref{eq:nnoiseflowU1U2}, are plotted in Fig.~\ref{fig:U1U2flows} for $d=2$.
At this order of perturbation, the fixed points of the RG equations for both conserved and nonconserved noise take the shape of hyperbolas (red curves) whose asymptotes (blue lines) mark transitions between two different behaviors. The red arrows indicate RG flows toward the stable fixed-point curves and, hence, show regions in the parameter space whose macroscopic behavior at criticality is described by the scaling exponents that we discuss in the next section (see Appendix~\ref{app_rgflowsdim} for the stability analysis of the fixed-point curves).
{
\color{black} The blue arrows, on the other hand, show runaway flows away from the Gaussian fixed point (red central dot). This runaway behavior may be the indication of the existence of a strong-coupling regime that cannot be captured by the one-loop perturbative approach, the signature of an inadequate starting point for the RG analysis where Gaussian power counting no longer applies, or, alternatively, the existence of a first-order phase transition~\cite{tauber2014}.
}

%
%
%

We have also analysed the RG flows in various spatial dimensions, with the results summarized in Fig.~\ref{fig:rgflowvardim}. Examining these flow equations (see Appendix~\ref{app_rgflowsdim}) shows that there are stable fixed points located along the $U_2$-axis (i.e. with $U_1=0$) for all $d$. On the contrary, fixed points on the $U_1$-axis (i.e. with $U_2=0$) only exist in $d=1$ in the case of conserved noise, and in $d \lesssim 2.27$ in the case of a nonconserved noise (such that fixed-point solutions for Eq.~\eqref{eq:nnoiseflowU1U2} with $U_2=0$ are available).
{\color{black}
The existence of the stable fixed-points on the $U_1$-axis with finite values of $\mu_1$ refers to macroscopic states with scaling behavior that are not usually considered in the context of KS systems, since the KS chemotactic interaction on its own is observed to lead to the formation of singular solutions \cite{hillen2009,chavanis2007}. This highlights the role of the noise in determining the macroscopic properties of an interacting system.
}
In $d=1$, one finds that the hyperbolas become fully attractive parallel straight lines and hence generic scaling behavior is expected throughout the $U_1$-$U_2$ space. Note that $d=1$ is a special case as the KS and 
polarity-induced chemotactic interactions ($\mu_1$ and $\mu_2$ in Eq.~\eqref{eq:langevin}) become proportional to each other. 
Furthermore, the two hyperbolas of fixed points become straight lines---coinciding with their asymptotes---at the upper critical dimension $d_c$. For $d>d_c$, the Gaussian fixed-point is stable, with the hyperbolas of fixed-point marking possible phase-transitions to strong coupling regimes. It should be stressed that in contrast to the exact scaling exponents, the RG flows discussed here are only valid up to one-loop in a perturbative expansion around the upper critical dimension, while higher-order terms may be needed in order to complete the picture of the flow diagram. 

\subsection{Exact scaling exponents}	\label{sec:expo}
\color{black}

We now focus on systems whose microscopic values of $U_{1,2}$ lie in the basin of attraction of the lines of fixed points described in the previous section. The scaling behavior of these systems is characterized by the critical exponents $\chi$ and $z$. For instance, for the long-time and large-scale particle density correlations one has the scaling form~\cite{tauber2014,medina1989}
\begin{align}
\left\langle \rho(\mathbf{x},t) \rho(\mathbf{x}',t') \right\rangle \sim |\mathbf{x}-\mathbf{x}'|^{2\chi} \, F\left( \frac{|t-t'|}{|\mathbf{x}-\mathbf{x}'|^z} \right), \label{eq:rhorhoscalingform}
\end{align}
where $F$ is the scaling function, and exponents $\chi$ and $z$ correspond to their (critical) values on the lines of fixed points. 

\charlie{
The first exponent identity, which follows from the Galilean symmetry and the non-renormalization of $(\mu_1-2\mu_2)$ and is valid at all orders of the perturbative expansion, reads:
%
%
\begin{align} \label{eq:exponidentity}
z + \chi = 0 \, .
\end{align}
This exact identity can also be checked at one-loop and directly follows from Eq.~\eqref{eq:cnoiseflowmu12} (or Eq.~\eqref{eq:nnoiseflowmu12} in the nonconserved noise case). 
}
%
%
The second exponent identity is a result of the nonrenormalization of the noise, as was discussed in Sec.~\ref{sec_RG}. In the case of a conserved noise, the identity reads as $z^{\rm con}-2\chi^{\rm con}=2+d$, which is obtained by setting $\partial_\ell \mathcal{D}_2 = 0$ in Eq.~\eqref{eq:cnoiseflowD2}. These relationships yield the exact exponents for $d<d^{\rm con}_c=4$:
\color{black}
\begin{align}
\label{eq:cnoiseexpons}
z^{\rm con} = - \chi^{\rm con} = (d+2)/3,
\end{align} 
in the case of conserved noise. 
A similar analysis for the nonconserved noise shows that for $d<d^{\rm non}_c=6$, the exact values of scaling exponents are
\begin{align}\label{eq:nnoiseexpons}
z^{\rm non}=-\chi^{\rm non}= d/3.
\end{align}
%
%
\color{black}
As a consequence of the Galilean symmetry and the nonrenormalization of the noise term, these critical exponents are \textit{exact}. Note that in both conserved and nonconserved noise cases, the exact exponents obtained are considerably different from their mean-field values, an indication of the importance of the fluctuations, especially close to a critical state.

\begin{figure}[t]
	\centering
		\includegraphics[width=0.48\textwidth]{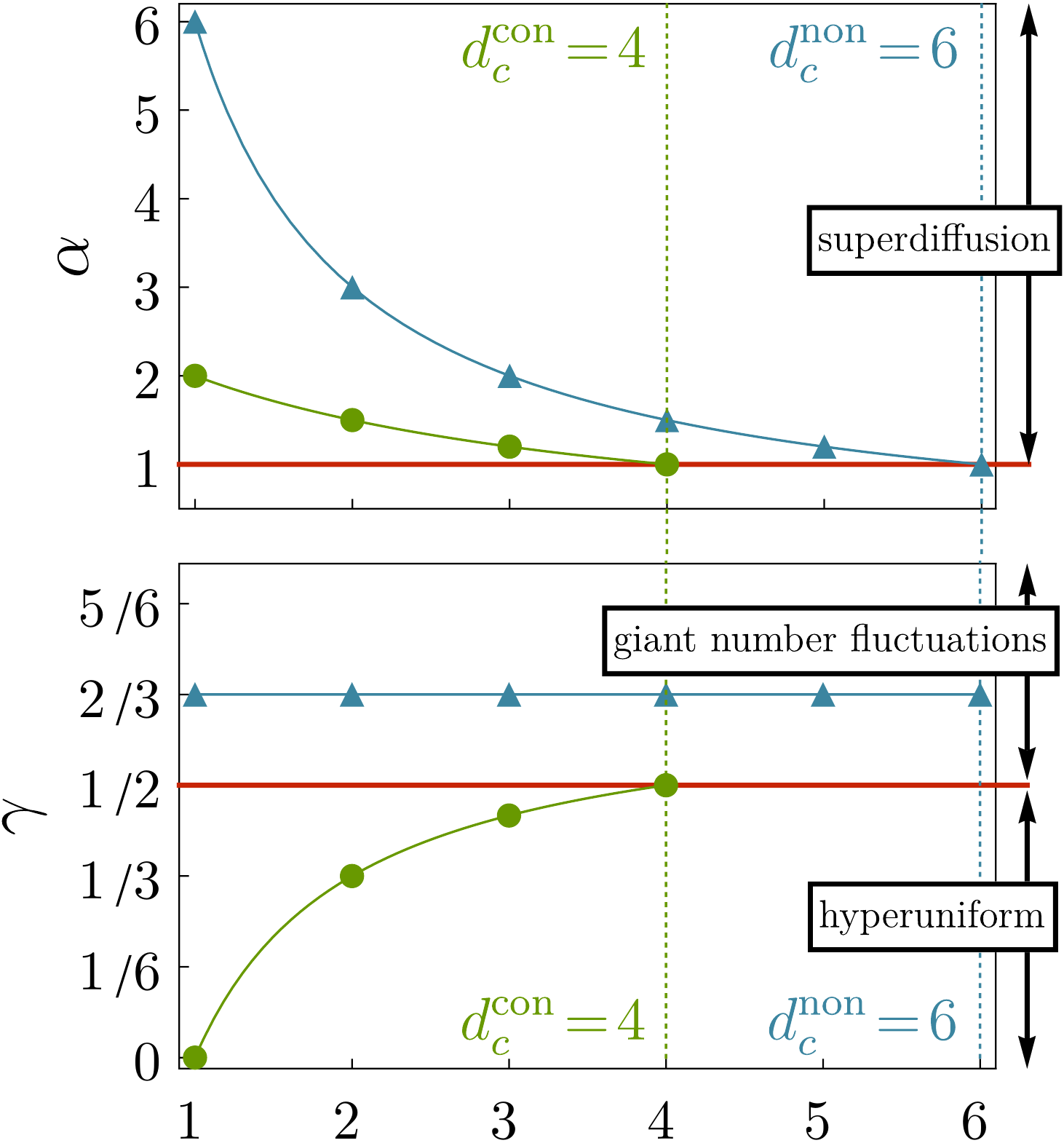}
	\caption{ Exact exponents $\alpha$ (top) and  $\gamma$ (bottom), which characterize the anomalous diffusion and number fluctuations (Eqs. \eqref{eq:alphaexponent} and  \eqref{eq:gammaexponent}) as functions of the dimension $d$, in the case of conserved (green circles) and nonconserved (blue triangles) noise.}
	\label{fig_exponents}
\end{figure}

To make a comparison with the case of simple diffusion, it is convenient to introduce the exponent $\alpha$ that characterizes how the mean-squared displacement depends on time, namely, $\Delta L^2 \equiv\left\langle \mathbf{x}(t)^2 \right\rangle \sim t^\alpha$ where 
\begin{align} \label{eq:alphaexponent-def}
    \alpha=\frac{2}{z}\,.
\end{align}
Note that $\alpha =1$ for diffusion. With the chemotactic interactions, on the other hand, we have:
\begin{align} \label{eq:alphaexponent}
    \alpha=
    \begin{cases}
    6/(d+2) \quad &\textrm{for conserved noise}, \\
     6/d \quad &\textrm{for nonconserved noise}.
    \end{cases}
\end{align}
\color{black}
For both conserved and nonconserved noise, one has $\alpha > \alpha_0=1$  as shown in Fig.~\ref{fig_exponents}, 
indicating that the chemotactic interactions result in \textit{superdiffusion} of the density fluctuations in the colony. 
In fact, in the case of nonconserved noise in $d<3$ we have $\alpha>2$, indicating an accelerated propagation of the density fluctuations. 
This accelerated propagation can be understood as a consequence of the long-range nature of the chemotactic interaction: fluctuations in the density due to the nonconserved noise influence the dynamics of the whole system and can lead to such a rapid propagation. 
This is not allowed in the conserved case where a noise-driven fluctuation is suppressed locally due to particle conservation. 
Note that the chemical field of each particle (governed by Eq.~\eqref{eq:poisson}) decreases faster with distance in higher dimensions, resulting in a reduction in the exponent $\alpha$ with dimension. 
As Eq.~\eqref{eq:rhorhoscalingform} implies, one can obtain this dynamic exponent in practice by measuring the spatial spreading of the density correlations in time.

%
%
%
%
%
%
%
%

We can also probe the statistics of the fluctuations in the number $N$ of particles within a subregion of volume $V$. While on average we have $\bar N=C_0 V$, the scale of fluctuations in $N$ defined via $\Delta N^2\equiv\left\langle (N-\bar N)^2 \right\rangle $ is influenced by the anomalous dimension of the density fluctuations as $\Delta N \sim \rho V$. This yields $\Delta N \sim \bar N^\gamma$ where 
\begin{align} \label{eq:gammaexponent-def}
    \gamma=1+\frac{\chi}{d}\,.
\end{align}
Note that $\gamma=1/2$ corresponds to Poissonian fluctuations. In the presence of chemotaxis, this exponent is given by
\begin{align} \label{eq:gammaexponent}
    \gamma=
    \begin{cases}
    (2/3)(1-1/d) \quad &\textrm{for conserved noise}, \\
     \, 2/3 \quad &\textrm{for nonconserved noise}.
    \end{cases}
\end{align}
For the conserved noise, $\gamma<\gamma^{\rm con}_0=1/2$ denotes a \textit{hyperuniform} density distribution. 
For the nonconserved noise, one has $\gamma = \gamma^{\rm non} =2/3 <\gamma^{\rm non}_0=1/2+1/d$ indicating \textit{giant number fluctuations} since $\gamma^{\rm non}>1/2$. 
Note that in both cases of conserved and nonconserved noise, the chemotactic interactions have resulted in a reduced value of the exponent $\gamma$ compared to its Gaussian value. 
Notably, the number fluctuations in the nonconserved case appear to be {\em superuniversal}, as the exponent $\gamma$ does not depend on the dimensionality of space. As the definition suggests, in experimental setups or simulations the exponent $\gamma$ can be calculated by measuring the extent of population fluctuations in subregions of the system.

%
%


\subsection{Emergence of the Galilean symmetry} \label{sec:symmetry}

The RG analysis and the exact critical exponents that we have derived in this section are a direct consequence of the Galilean symmetry of Eq.~\eqref{eq:langevin} governing the dynamics of the density fluctuations. 
However, note that neither the microscopic dynamics in Eq~\eqref{eq:cell_velocity} nor the full DK equation \eqref{eq:continuity_DK} is actually invariant under the symmetry transformations~\eqref{eq:galileantransform} and the Galilean symmetry only \textit{emerges} at the macroscopic scale upon expanding the density in the DK equation (see Section.~\ref{sec:langevin}). 
By assuming that the density fluctuations $\rho$ are sufficiently small compared to the average density $C_0$ (i.e., $\rho \ll C_0$),  one discards the symmetry-breaking term
$\nabla\cdot \left(\mu_3\rho\nabla(\nabla\phi)^2\right)$ 
from the expansion (which would appear in the Langevin dynamics~\eqref{eq:langevin} if the full particle current~\eqref{eq:JDK_fullexpansion} is kept) since this term scales as $\rho^3$ and its coupling $\mu_3$ is also irrelevant under RG (its engineering dimension is $[\mu_3]_0=-2$ at the upper critical dimension; see Appendix~\ref{app_powerCounting}).
We note that this is a common practice in constructing hydrodynamic equations using a gradient expansion approach to coarse-grain the microscopic dynamics.

From an RG perspective, the presence of the symmetry-breaking term $\propto \mu_3$---albeit irrelevant---has the potential to change the structure of the RG flows, as it may generate \textit{relevant} terms that break the Galilean symmetry, if such terms exist. 
In the case of the Langevin equation~\eqref{eq:langevin}, a power-counting analysis shows that the only relevant symmetry-breaking term is 
$ \nabla \cdot \left( \mu_4 (\nabla\phi)^2 \nabla\phi \right)$ 
(see Appendix~\ref{app_powerCounting}).
Accordingly, this $\propto\mu_4$ term cannot be safely discarded from the macroscopic theory unless an additional symmetry shared by the Galilean-invariant $\mu_1$, $\mu_2$, and $\mu_3$ terms prevents its generation along the RG flow.
We have been able to identify such a symmetry which can be characterized by the following transformation of the chemical gradient:
\begin{align}   \label{eq:crosstransf}
    \nabla \phi \to \nabla \phi + \varepsilon \nabla f(\rho) \times \nabla \phi,
\end{align}
where $\varepsilon$ is the transformation parameter, $f$ is an arbitrary function of $\rho$ with $\nabla f(\rho)= f'(\rho)\nabla\rho\neq0$, and $\times$ represent the vectorial cross product.
We note that  the Poisson equation~\eqref{eq:poisson} is invariant under this transformation since 
$\nabla\cdot(\nabla f(\rho) \times \nabla\phi) = \nabla\cdot \left( \nabla\times ( f(\rho) \nabla\phi) \right) = 0$. 
On the other hand, the full DK current~\eqref{eq:JDK_fullexpansion} (which includes also the $\mu_3$ contribution), from which the r.h.s of the Langevin equations~\eqref{eq:langevin} is derived, transforms as
\begin{align}   \label{eq:JDKepsilontransf}
    \nabla \cdot \mathbf{J}_{\rm DK} \to
    \nabla \cdot \mathbf{J}_{\rm DK} +  \mathcal{O}(\varepsilon^2),
\end{align}
while the possible additional current $\mathbf{J}_4 = \mu_4 (\nabla\phi)^2 \nabla\phi$ due to the relevant term $\propto \mu_4$ changes as
\begin{align}   \label{eq:J4epsilontransf}
    \nabla\cdot \mathbf{J}_4
    \to
    \nabla\cdot \mathbf{J}_4
    + 
    \varepsilon \nabla \cdot \left( \mu_4 (\nabla\phi)^2 \nabla f(\rho) \times \nabla\phi \right) + \mathcal{O}(\varepsilon^2),
\end{align}
Equations~\eqref{eq:JDKepsilontransf} and \eqref{eq:J4epsilontransf} show that in the limit 
$\varepsilon \to 0$, the transformation~\eqref{eq:crosstransf} is an infinitesimal symmetry of the Langevin dynamics with $\mu_1$, $\mu_2$, and $\mu_3$ couplings, whereas the $\mu_4$ term breaks this symmetry. 

We note that this additional symmetry is related to the fact that in the microscopic dynamics~\eqref{eq:cell_velocity}, both $\nu_1$ and $\nu_2$ terms are gradients of the corresponding chemical potentials 
$\nu_1 \Phi$ and $\frac{\nu \chi}{6D_r}(\nabla\Phi)^2$, and therefore they represent the effects of irrotational force fields.  
On the contrary, the microscopic force $\nabla\Phi (\nabla\Phi)^2$ which eventually gives rise to the $\mu_4$ term (see Appendix~\ref{app:moment_long}) is not the gradient of any function and therefore has nonzero vorticity. 
In principle, the transformation~\eqref{eq:crosstransf} excludes the possibility of generating rotational force fields from coarse-graining irrotational forces.
We further corroborated this argument based on the additional symmetry by computing explicitly the one-loop RG flows of the couplings in the presence of the irrelevant $\mu_3$ term, which does not turn out to generate the relevant $\mu_4$ coupling at this order. 

%


The identification of emergent macroscopic symmetries is of crucial importance: in high-energy physics for instance, it has led to the modern understanding of symmetries which were presumed to be fundamental~\cite{witten2018}, such as baryon and lepton number conservation~\cite{weinberg1979}. These symmetries, rather than being fundamental, can in fact be seen as low-energy accidents, emerging as a consequence of gauge symmetries. They emerge because the only gauge-invariant operators that one can construct within the standard model yield negligible contributions at classical energy levels, although they are important at higher energy.
%
In our case, the irrelevant non-Galilean invariant $\mu_3$ term, despite being present at the microscopic level, flows towards vanishing values upon iterations of the RG transformation. Since it cannot generate the relevant $\mu_4$ term, the Galilean symmetry 
emerges at the macroscopic level and leads to the exact scaling exponents obtained in this section.

\color{black}

\section{Concluding remarks}	\label{sec:conc}

\color{black}
In this work, we have introduced a novel mechanism for chemotaxis induced by the polarity response of the particles and have investigated how it affects the collective macroscopic dynamical properties of the system.

At the microscopic level, the 
polarity-induced mechanism that we have studied is expected to arise when the cell can undergo a 
polarity change---achieved through shape changes or redistribution of surface receptors---in response to an external chemical gradient~\cite{roussos2011,iglesias2008}. 
This type of response is known to be prevalent in eukaryotic cells~\cite{iglesias2008,levine2013}, and reported in the context of chemotactic response of chemically active colloids~\cite{saha2014} and enzymes~\cite{adeleke-larodo2019}. 
A manifestation of this response can arise in bacteria as well due to a coupling between the asymmetric geometry and the spatial distribution of sensors~\cite{kranz2016,gelimson2016}.

Starting from the microscopic equations, we have derived a mesoscopic mean-field description of these particles by averaging over the fast, orientational, degrees of freedom,
which upon implementing the noise term gives the DK equation~\eqref{eq:continuity_DK} for the full particle density.
Focusing on the limit of fast diffusion and slow degradation of the chemical signals, which means the chemical field fluctuations adapt immediately to the fluctuations of the particle density and obey a Poisson equation~\eqref{eq:poisson}, we then obtain the Langevin equation \eqref{eq:langevin} for the particle density fluctuations by expanding the DK equation around a uniform density $C_0$.
In the resulting coarse-grained description, the 
polarity-induced chemotactic mechanism $\mu_2 \nabla^2(\nabla\phi)^2$ appears to be equally relevant as the
KS term $\mu_1 \nabla\cdot(\rho \nabla\phi)$. 
Since $\mu_2$, as opposed to $\mu_1$, is proportional to the mean particle density $C_0$, we understand that this relevant interaction becomes stronger in systems with dense populations.
We also show that, contrary to the KS term, the 
polarity-induced interaction cannot be derived from a functional, and hence represents a genuine nonequilibrium term.

We demonstrate that the Langevin equation~\eqref{eq:langevin} is invariant under the Galilean transformation given by Eq.~\eqref{eq:galileantransform}. 
Although broken at the microscopic level by the presence of an irrelevant symmetry-breaking term, this symmetry emerges at larger scales (see discussion in Sec.~\ref{sec:symmetry}).%
It is worth mentioning that equipped with this symmetry, Eq.~\eqref{eq:langevin} could also be directly derived from a systematic expansion in $\rho$ and $\phi$ by including all the relevant Galilean-symmetric terms, and can thus be seen as the natural extension of the KS model preserving Galilean symmetry. 

\color{black}
As a result of this emergent Galilean symmetry, the chemotactic couplings $\mu_{1,2}$ are not affected by the RG flow, providing an exact exponent identity (see Eq.~\eqref{eq:exponidentity}). With the nonrenormalization of the noise strength, 
these findings enable us to obtain the dynamical scaling exponents exactly whose values indicate superdiffusive  
propagation of density fluctuations  
with non-Poissonian distributions, either in the form of hyperuniform populations (conserved noise) or exhibiting giant number fluctuations (nonconserved noise), see Eqs.~\eqref{eq:cnoiseexpons} and \eqref{eq:nnoiseexpons} as well as Fig.~\ref{fig_exponents}. 
The fixed points of the RG flows for the effective chemotactic couplings $U_{1,2}$ that (unlike the exact exponents) are only one-loop results, represent a pair of hyperbolas with identical scaling exponents throughout (see Fig.~\ref{fig:U1U2flows} and Appendix~\ref{app_rgflowsdim} for details). 
The scaling behavior described here is particularly relevant for polarizable particles and can be searched for by measuring the scaling exponents, as outlined by Eqs.~\eqref{eq:alphaexponent} and \eqref{eq:gammaexponent} and the discussion thereof.

The interplay between chemical signals and generic growth processes of the particles, which in many cases are asymmetrical processes accompanied by the 
polarity of the cells \cite{jan2000,neumuller2009}, adds another level of complexity to the collective properties of growing colonies  \cite{kruse2005,toner2012a,gelimson2015,malmi-kakkada2018} which we plan to investigate in future works. Although the growth of individuals is known to be limited by conditions such as the availability of nutrients in an environment \cite{wang2017} and cell homeostatic regulations \cite{tzur2009}, the complex internal machinery determining the size and dynamic structure of the colony remains largely unknown. Such self-regulations are crucial in the development of different organs in the body and show signs of failure when, for instance, tumor cells acquire increased proliferation by breaking away from these self-regulations \cite{preston-martin1990, hanahan2011}.
Input from powerful physical considerations such as scaling properties and symmetry transformations are crucial for choosing the most relevant interactions from a large number of possibilities that could be included in theoretical models.
An understanding of different phases of the system in the presence of both chemical signals and growth processes will help us to identify such regulatory mechanisms.


\acknowledgements

S.M. thanks T.~Adeleke-Larodo for helpful discussions on moment expansion.
R.B.A.Z. and C.D. thank B. Delamotte for arousing their curiosity about this model, and for stimulating discussions.  
S.M. is supported by a joint Clarendon and Kendrew Scholarship
from the University of Oxford and St John's College. 
A.G. acknowledges support from the MIUR PRIN project ``Coarse-grained description for non-equilibrium systems and transport phenomena (CO-NEST)'' n. 201798CZL.
R.G. acknowledges support by the MaxSynBio Consortium which is jointly funded by the Federal Ministry of Education and Re-
search of Germany and the Max Planck Society.

\appendix


\color{black}
    
\section{Toy models}    \label{app_toymodel}

In this section, we discuss two simple yet representative toy models: the first one considers a chemotactic particle with an arbitrary distribution of gradient-sensing units on its surface and is a natural extension of a similar earlier work~\cite{kranz2016,gelimson2016}. The second one, instead, considers a basic cell composed of two chemotactic force-generating units that can be spatially separated by the presence of a chemical gradient. Both models illustrate how, within these simplified descriptions, the 
polarity-induced chemotaxis emerges, in addition to the usual gradient-sensing KS chemotaxis.
It should be emphasized that these models are not meant to capture all possible biological or chemical mechanisms that lead to the generalized chemotaxis studied in this paper. They are introduced to show how this novel mechanism for chemotaxis may naturally result from relatively basic extensions of what leads to the KS term, and provide a conceptual framework for similar derivations in other systems.

\begin{figure}[t]
	\centering
		\includegraphics[width=0.45\textwidth]{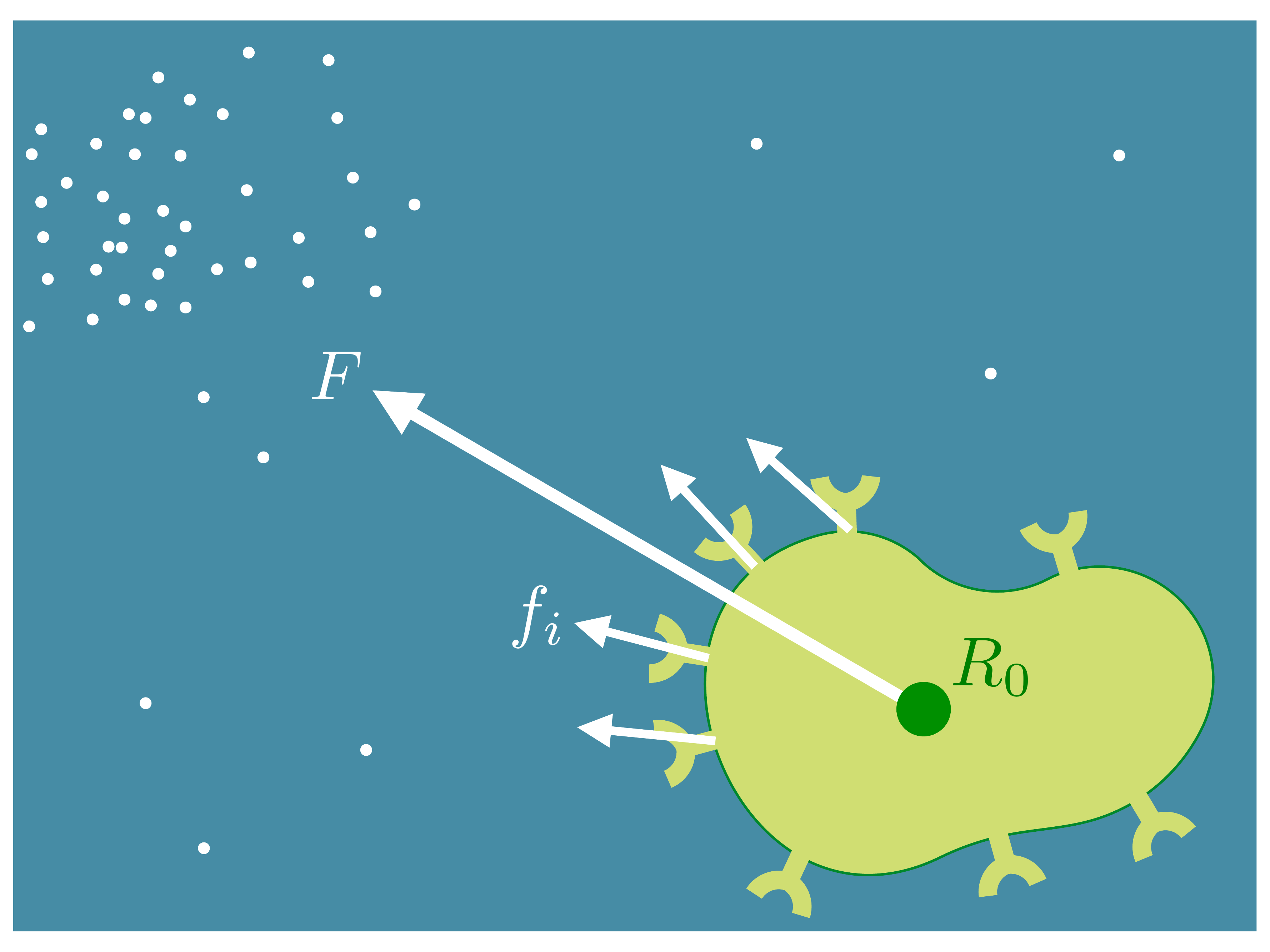}
	\caption{Schematic of a chemotactic cell (green blob) in a chemical field (white dots). Each sensing unit exerts a force $\bm f_i$ that adds up to give the resultant $\mathbf{F}$; see Eq.~\eqref{eq:F_gradient_sensing}.}
	\label{fig:toy_model_2}
\end{figure}

\subsection{Polar particles}

Consider a set of $M$ gradient-sensing units, distributed on the surface of a cell  (see Fig.~\ref{fig:toy_model_2}). A single unit, located at position $\mathbf{r}_i$, is assumed to exert a force 
\begin{align}
    \mathbf{f}_i = \Upsilon_i \nabla \Phi (\mathbf{r}_i) \, , \label{eq_gradientSensing}
\end{align}
where the chemotactic coupling strength $\Upsilon_i$ may depend on the unit. We note that this gradient-sensing mechanism can, for instance, stem from a spatial coarse-graining of smaller subunits sensing the (absolute) value of the chemical concentration $\Phi$ or, alternatively, can originate from a temporal coarse-graining where each unit locally estimates and responds to \mbox{$\Phi(\mathbf{r}_i(t+\delta t))-\Phi(\mathbf{r}_i(t))$} during the time scale of $\delta t$.

Let us define
$\mathbf{R}_0 =\frac{1}{M} \sum_{i=1}^M  \mathbf{r}_i$ 
as the centroid of the cell, and 
$\delta \mathbf{r}_i=\mathbf{r}_i-\mathbf{R}_0$.
The total force exerted by all the units can be expanded around $\mathbf{R}_0$ as:
\begin{align} \label{eq:F_gradient_sensing}
    \mathbf{F} &= \sum_{i=1}^M \mathbf{f}_i 
    = \sum_i \Upsilon_i \nabla  \Phi (\mathbf{R}_0 + \delta \mathbf{r}_i ) \\
    &= \sum_{i=1}^M \Upsilon_i \Big[ \nabla  \Phi (\mathbf{R}_0) +  \delta \mathbf{r}_i \cdot \nabla \nabla \Phi (\mathbf{R}_0) + {\mathcal O}(\delta \mathbf{r}_i^2) \Big]  \, .
\end{align}
Balancing this force against a frictional force $-\Xi \mathbf{v}$ due to motion of the whole particle with velocity $\mathbf{v}$ and where $\Xi$ is an effective translational friction coefficient, we can find an expression for the translational velocity of the form given in Eq.~\eqref{eq:v-nu1-nu2-1}, namely
\begin{align}
    \mathbf{v} = \nu_1 \nabla  \Phi (\mathbf{R}_0) + \nu_2  \mathbf{n} \cdot \nabla \nabla \Phi (\mathbf{R}_0) + \cdots \, ,
\end{align}
where
\begin{align}
\nu_1 = \frac{1}{\Xi}\sum_{i=1}^M \Upsilon_i \, ,
\end{align}
and 
\begin{align}
\nu_2 = \frac{1}{\Xi} \biggl| \biggl|  \sum_{i=1}^M \Upsilon_i \delta \mathbf{r}_i \biggr| \biggr| \, ,
\end{align}
define the effective coupling constants, and
\begin{align}
\mathbf{n} = \frac{\sum_{i=1}^M \Upsilon_i \delta \mathbf{r}_i}{ \biggl| \biggl|  \sum_{i=1}^M \Upsilon_i \delta \mathbf{r}_i \biggr| \biggr|} \, ,
\end{align}
defines the polarity of the cell.

The forces $\mathbf{f}_i$ also exert a net torque $\bm \tau$ on the particle which tends to rotate the polarity $\mathbf{n}$ and can be calculated as
\begin{align}
    \bm \tau &= \sum_{i=1}^M  \delta \mathbf{r}_i \times \mathbf{f}_i
    = \sum_{i=1}^M \Upsilon_i \delta \mathbf{r}_i \times \nabla \Phi(\mathbf{R}_0 + \delta \mathbf{r}_i) \\
    &= \sum_{i=1}^M \Upsilon_i \delta \mathbf{r}_i \times \Big[ \nabla \Phi(\mathbf{R}_0)+ {\mathcal O}(\delta \mathbf{r}_i) \Big]\, .
\end{align}
Balancing this torque in the overdamped regime against a frictional torque $-\Xi_r \bm \omega$ due to rotation with angular frequency $\bm \omega$, where $\Xi_r$ is an effective rotational friction coefficient, we can find an expression for the rotational velocity of the form given in Eq.~\eqref{eq:align}, namely
\begin{equation}
    \bm \omega= \chi \mathbf{n} \times \nabla \Phi (\mathbf{R}_0) + \cdots \, ,
\end{equation}
where we have introduced
\begin{align}
\chi = \frac{1}{\Xi_r}\biggl| \biggl|  \sum_{i=1}^M \Upsilon_i \delta \mathbf{r}_i \biggr| \biggr| \, .
\end{align}

    \subsection{Extensible particles} \label{app_toy_extens}

\begin{figure}[t]
	\centering
		\includegraphics[width=0.45\textwidth]{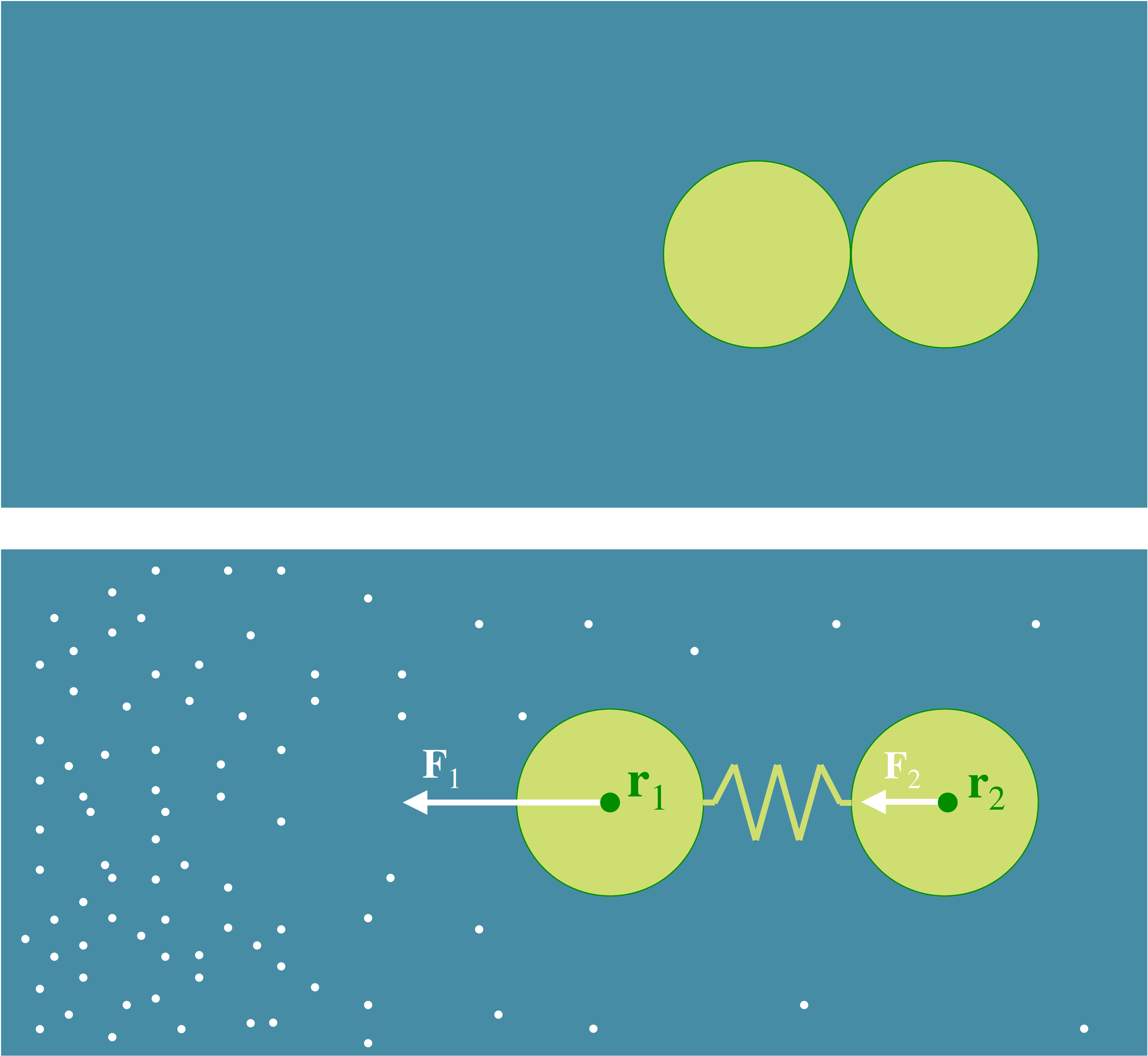}
	\caption{Schematic of an extensible particle in uniform chemical background (upper row) and in the presence of a chemical gradient (lower row).}
	\label{fig:twobeads}
\end{figure}

For this second toy model, more basic than the first one, we consider a scenario where a particle consists of two force-generating units, as sketched in Fig.~\ref{fig:twobeads}. In presence of a chemical gradient, we assume that these two units can get separated in space, such that the cell is polarized and its equation of motion can be cast in the form of Eq.~\eqref{eq:cell_velocity}, as we show below.

The units are assumed to generate forces which are proportional to the chemical gradient at their location. %
The total force $\mathbf{F}$ generated by the two units located at positions $\mathbf{r}_{1,2}$ can be obtained as
\begin{align}
    \mathbf{F} &= \Upsilon \nabla \Phi ( \mathbf{r}_1 ) + \Upsilon \nabla \Phi (\mathbf{r}_2 ), \\
    &= 2\Upsilon \nabla \Phi(\mathbf{r}_1) 
    + \Upsilon  \delta\bm{l}\cdot\nabla\nabla \Phi(\mathbf{r}_1) 
    + \mathcal{O}(\delta\bm{l}^2) 
\end{align}
where $\Upsilon$ is the chemotactic couplings determining the strength of the forces generated by the two units. 
In writing the second line, we have Taylor expanded the force generated by unit $2$ by defining 
$\delta \bm{l} = \mathbf{r}_2 - \mathbf{r}_1$
and we have discarded the second-order terms in the expansion, 
which is justified as far as the chemical gradient does not change appreciably across the length of a particle. 

We further assume that $\delta\bm{l} =  k \nabla \Phi$, meaning that the force-generating units are actually separated in space due to the chemical gradient. 
This indicates that the arrangement of the units on the particle or, alternatively, the particle's shape, is affected by the chemical gradient and we have retained the linear approximation of such an effect. 
Eventually, we note that in the overdamped regime for the motion of the particle in the surrounding medium, the frictional force due to the velocity $\mathbf{v}$ of the whole particle in the form of $-\Xi \mathbf{v}$, where $\Xi$ is the friction coefficient, balances the total force generated by the units which compose it. 
We can therefore obtain an expression for the translational velocity similar to Eq.~\eqref{eq:cell_velocity} which reads
\begin{align}
    \frac{ {\rm d} \mathbf{r}}{ {\rm d} t} = 
    \frac{2 \Upsilon}{\Xi} \nabla \Phi  
    + \frac{ k \Upsilon}{2 \Xi}  \nabla\left(\nabla\Phi\right)^2.
\end{align}
We note that in this more primitive model, the coefficient in front of $\nabla \left(\nabla\Phi\right)^2$ stems from the induced polarity of the particle in response to the chemical gradient, as opposed to the averaging of the polarity dynamics in the previous toy model.


\section{Derivation of Equation~\eqref{eq:n_handwaving} for average polarity} \label{app:moment}

In this section we detail the averaging over the polarity degrees of freedom that was outlined above. For simplicity, we consider a $3$-dimensional system in this section, as the generalization to other dimensions is straightforward. 

Consider a collection of $N$ polar particles with positions $\mathbf{r}_a(t)$ and polarity unit vectors $\mathbf{n}_a(t)$. Building on the microscopic equations that govern the individual particle dynamics, we derive a Fokker--Planck equation for the probability distribution $\mathcal{P}$ of position $\mathbf{x}$ and 
polarity $\mathbf{n}$ of the particles, defined by 
\begin{align}       \label{eq:Prob_meanfield}
\mathcal{P}(\mathbf{x}, \mathbf{n};t) = \left\langle \sum_{a=1}^N \delta\left(\mathbf{x}-\mathbf{r}_a(t)\right)\delta\left(\mathbf{n}-\mathbf{n}_a(t)\right) \right\rangle,
\end{align}
where the average is over all different realizations of the system. 
The Langevin equations for the position and polarity of the individual particles read as
\begin{align}
    \frac{\mathrm{d}}{\mathrm{d}t}\mathbf{r}_a(t) &= 
    \bm{v}_{\rm KS} (\mathbf{r}_a) + \bm{v}_{\rm p} (\mathbf{r}_a,\mathbf{n}_a) +\bm{\xi}_a(t),   
    \label{eq:microTLangevin} \\
    \frac{\mathrm{d}}{\mathrm{d}t}\mathbf{n}_a(t) &= 
    \chi \mathbf{n}_a \times \nabla \Phi (\mathbf{r}_a)+ \bm{\gamma}_a(t)\times\mathbf{n}_a, 
    \label{eq:microRLangevin}
\end{align}
where 
we have used Eqs.~\eqref{eq:v-nu1-nu2-1} and \eqref{eq:align} for the deterministic parts of the translational and angular velocities experienced by 
particles.
Here, $\bm{\xi}_a$ and $\bm{\gamma}_a$ are Gaussian white noise terms acting on the $a$th particle, characterized by $ \langle\bm\xi_a(t)\rangle = 0$, and
\begin{align}   \label{eq:microTnoise} 
\langle \xi_{al}(t) \xi_{bm}(t') \rangle = 
2D\delta_{ab}\delta_{lm} \delta(t-t'), 
\end{align}
as well as, $\langle\bm\gamma_a(t)\rangle = 0$, and
\begin{align} \label{eq:microRnoise}
\langle \gamma_{al}(t) \gamma_{bm}(t') \rangle = 
2 D_r \delta_{ab}\delta_{lm}\delta(t-t'). 
\end{align}
Here $l$ and $m$ represent different components, $D$ is the translational diffusion coefficient, and 
$D_r$
is the effective re-orientation rate as biased by the gradient (see Sec. \ref{sec:polareffects} and Ref.~\cite{schnitzer1993}). Note that we assume the stochastic forces acting on different particles to be uncorrelated.

The time evolution of $\mathcal{P}(\mathbf{x},\mathbf{n};t)$ can be obtained in the form of a Fokker--Planck equation
\begin{equation} \label{eq:FP}
\begin{split}
\partial_t\mathcal{P} = 
&- \nabla\cdot\left[ 
\left(\bm{v}_{\rm KS} + \bm{v}_{\rm p} \right)\mathcal{P} - D\nabla\mathcal{P}\right] \\
&- \bm{\mathcal{R}} \cdot 
\left[ \left(\chi \mathbf{n} \times \nabla \Phi \right) \mathcal{P} - D_r \bm{\mathcal{R}}  \mathcal{P} \right], 
\end{split}
\end{equation}
where $\bm{\mathcal{R}}=\mathbf{n}\times\nabla_{\mathbf{n}}$ is the rotational gradient operator~\cite{doi1988}. 
We note that this Fokker--Planck equation is at the mean-field level and neglects the statistical correlations between the particles (which are taken into account in Sec. \ref{sec:stoch}).

A hierarchy of equations can now be constructed starting from Eq.~\eqref{eq:FP} by considering various moments of the distribution $\mathcal{P}$ with respect to $\mathbf{n}$. 
For instance, the particle density is defined as 
\begin{align}   \label{eq:C_meanfielddef}
C (\mathbf{x},t) = \int_{\mathbf{n}} \mathcal{P}(\mathbf{x},\mathbf{n};t) 
= \left\langle \sum_{a} \delta(\mathbf{x}-\mathbf{r}_a(t)) \right\rangle,
\end{align}
while the polarization field is given by
\begin{align}   \label{eq:p_meanfielddef}
\mathbf{p}(\mathbf{x},t) = \int_{\mathbf{n}} \mathbf{n} \mathcal{P}(\mathbf{x},\mathbf{n};t) 
= \left\langle \sum_a \mathbf{n}_a \delta(\mathbf{x}-\mathbf{r}_a(t)) \right\rangle.
\end{align}
The equation governing the density can be obtained by integrating Eq.~\eqref{eq:FP} with respect to the polarity $\mathbf{n}$ which leads to the continuity equation
\begin{align} \label{eq:continuity_meanfield}
    &\partial_t C (\mathbf{x},t) +\nabla\cdot 
    \mathbf{J} (\mathbf{x},t)=0,
\end{align}
where the particle current is given by 
\begin{equation} \label{eq:J_meanfielddef}
\begin{split}
\mathbf{J} (\mathbf{x},t) =
    -D\nabla C & +  
    \nu_1 C \nabla\Phi 
    + \nu_2 \mathbf{p} \cdot \nabla\nabla\Phi,
\end{split}
\end{equation}
which depends on $C$ but also on the polarization field $\bf{p}$.
The interdependence among the equations governing different moments is, in fact, a general feature of these hierarchical equations at all orders \cite{golestanian2019}. 
The dynamics of the polarization field can be obtained in similar fashion, by multiplying both sides of Eq.~\eqref{eq:FP} by $\mathbf{n}$ and performing an integration over $\mathbf{n}$ resulting in 
\begin{equation} \label{eq:polarizationFP}
\begin{split}
    \partial_t  p_i &+ 
    \partial_l \bigg[ 
    -D\partial_l p_i  
    +\nu_1  p_i  \partial_l\Phi 
    + \nu_2 \left( Q_{ik} + \frac{\delta_{ik}}{3} C \right)  
    \partial_k \partial_l \Phi
    \bigg] \\
    &+ 2 D_r p_i - \frac{2\chi}{3}  C \partial_i\Phi
    + \chi Q_{il}\partial_l \Phi = 0, 
\end{split}
\end{equation}
where $Q_{ij}$ represents the nematic order parameter tensor and is defined as $\bm{Q}(\mathbf{x},t) = \int_{\mathbf{n}} \left[\mathbf{n}\mathbf{n}-\frac{1}{3} \, \mathbf{I}\right]\mathcal{P}(\mathbf{x},\mathbf{n};t).$
The dynamics of the nematic order parameter field is then obtained as
\begin{equation}
\begin{split}
    & \partial_t Q_{ij} + \frac{1}{3}\partial_t C \delta_{ij}+6  D_r  Q_{ij} + 2\chi  Q_{ijl}^{(3)} \partial_l \Phi   \\
    &+ \partial_l \bigg[ 
    \nu_1 \partial_l\Phi\left(Q_{ij} + \frac{1}{3}C \delta_{ij}\right)  - D\partial_l\left(Q_{ij}+ \frac{1}{3}C \delta_{ij}\right)  \\
    &+ \nu_2 \, \partial_k\partial_l\Phi \left(Q^{(3)}_{ijl} + \frac{1}{5}\left(p_i\delta_{jk}+p_j\delta_{ik}+p_k\delta_{ij}\right)\right)\bigg]  \\
    &-\frac{\chi}{5} \bigg(3 \left(p_i\partial_j \Phi + p_j\partial_i \Phi\right)-2\delta_{ij} p_l \partial_l\Phi \bigg)= 0, 
\end{split}
\end{equation}
where $Q_{ijl}^{(3)}$ is the third-order moment and its expression is given in Eq.~\eqref{eq:Q3_def} of Appendix~\ref{app:moment_long}. At long time and for large length scale, one can take the hydrodynamic limit and only keep the lowest order in gradients. One observes from the previous equation that $Q_{ij}= \mathcal{O}(\nabla)$ in this limit, such that it can be discarded from Eq.~\eqref{eq:polarizationFP}, as well as all the derivatives of $p_i$. We thus obtain, in the hydrodynamic limit:
%
\begin{align}   \label{eq:p_1st}
    \mathbf{p} \approx \frac{\chi}{3 D_r} C \partial_i \Phi,
\end{align}
in agreement with Eq.~\eqref{eq:n_handwaving}. 
Substituting Eq.~\eqref{eq:p_1st} into Eq.~\eqref{eq:J_meanfielddef}, we get for the particle current
\begin{equation} \label{eq:J_final}
    J_i = 
    - D \partial_i C 
    + \nu_1 C \, \partial_i\Phi  
    + \frac{\nu_2\chi}{3D_r} C \,\partial_l\Phi \, \partial_l\partial_i\Phi. 
\end{equation}
Finally, we recover Eq.~\eqref{eq:cell_velocity} from this particle current, which concludes the moment expansion analysis.

    
\section{Detailed balance} \label{app_detailedbalance}
    

\color{black}

The aim of this appendix is to demonstrate that the novel 
polarity-induced chemotactic term cannot be derived from a free energy and is a purely nonequilibrium interaction, which breaks detailed balance.
We establish the validity of this point using a proof-by-contradiction approach, where assuming the existence of a free energy whose functional derivative gives the $\frac{\nu_2 \chi}{6 D_r}$ term leads to a contradiction in that the second order derivatives of the assumed free energy do not commute. 

As mentioned in the main text, the dynamics for the density field Eq.~\eqref{eq:continuity_DK} can be expressed as:
\begin{align}
    \partial_t \hat{C} =
    -\nabla \cdot \left[ - \hat{C} \nabla \frac{\delta \mathcal{F}_{\rm KS}}{\delta \hat{C} } + \frac{\nu_2 \chi}{6 D_r} \hat{C} \nabla (\nabla \hat{\Phi})^2 + \sqrt{\hat{C}} \,\bm \xi  \right],
    \label{eq:fullDKappendix}
\end{align}
where the KS functional $\mathcal{F}_{\rm KS} $ has been introduced in Eq.~\eqref{eq:KSfunctional}. 
In order to verify this form through functional differentiation, we take into account the fact that the particles are sources of the chemicals, and hence one can write the chemical field $\Phi$ as
\begin{align}   \label{eqC:phikernel}
    \hat{\Phi}(\mathbf{x}) = \int \mathrm{d}^d\mathbf{y} K(\mathbf{x}-\mathbf{y}) \hat{C} (\mathbf{y})
\end{align}
where the screened Coulomb kernel $K$ satisfies the condition 
$\left(-\nabla^2 + \kappa^2\right) K(\mathbf{x}-\mathbf{y})= \delta^d(\mathbf{x}-\mathbf{y})$ 
imposed by the screened Poisson equation, Eq.~\eqref{eq:Phi_exactPoisson}. Note that in analogy to electrostatics, $K(\mathbf{x}-\mathbf{y})$ is the potential at point $\mathbf{x}$ due to a unit source at point $\mathbf{y}$ and $-\nabla K(\mathbf{x}-\mathbf{y})$ gives the corresponding chemotactic drift at position $\mathbf{x}$, which is parallel to $\mathbf{x}-\mathbf{y}$.
The functional derivative of $\mathcal{F}_{\rm KS}$ can then be computed as 
\begin{align}
    \frac{\delta \mathcal{F}_{\rm KS} }{\delta \hat{C}(\mathbf{x}) }
    &=
    D ( {\rm log} \hat{C}(\mathbf{x}) + 1 ) \\
    &\qquad\quad
    -\frac{\nu_1}{2}
    \Bigg[
      {\hat \Phi} (\mathbf{x}) + \underbrace{\int \mathrm{d}^d \mathbf{y} \,  K( \mathbf{x}- \mathbf{y} ) \hat{C}(\mathbf{y})}_{\hat{\Phi}(\mathbf{x})} 
    \Bigg] ,\nonumber
\end{align}
yielding
\begin{align}
    \hat{C} \nabla \frac{\delta \mathcal{F}_{\rm KS} }{\delta \hat{C}} &=
    D\nabla \hat{C}
    -\nu_1 \hat{C} \nabla \hat{\Phi},
\end{align}
as required for the KS current. 
We conclude that the KS contribution to the particle current, taken separately, results from an equilibrium interaction thereby satisfying the detailed balance condition.

On the contrary, the $\nu_2$ term in the DK equation~\eqref{eq:fullDKappendix}, which gives rise to the $\mu_2$ interaction in Eq.~\eqref{eq:langevin}, cannot be derived from a free energy functional. In order to show this, consider a (hypothetical) free energy $\mathcal{F}_2$ whose functional derivative is assumed to give the $\frac{\nu_2 \chi}{6 D_r}$ interaction term, i.e. 
$\delta \mathcal{F}_2 / \delta \hat{C}(\mathbf{x})= \left(\nabla \hat{\Phi} (\mathbf{x})\right)^2$.
Below we show that the second derivatives of $\mathcal{F}_2$ do not commute, which amounts to the breakdown of the Onsager relations for equilibrium interactions.
To demonstrate this, let us take a derivative from the above expression, and obtain
\begin{align}
    \frac{\delta^2\mathcal{F}_2}{\delta \hat{C} (\mathbf{x}') \delta \hat{C}(\mathbf{x})} = 
    2 \left(\nabla \hat{\Phi} (\mathbf{x})\right)\cdot \frac{\delta \nabla \hat{\Phi}(\mathbf{x})}{\delta \hat{C}(\mathbf{x}')}.
\end{align}
Making use of Eq.~\eqref{eqC:phikernel} and interchanging $\mathbf{x}$ and $\mathbf{x}'$ then yields
\begin{align}
    \frac{\delta^2 \mathcal{F}_2}{\delta \hat{C} (\mathbf{x}') \delta \hat{C} (\mathbf{x})} 
    &= 2 \nabla\hat{\Phi}(\mathbf{x})\cdot\nabla K(\mathbf{x}-\mathbf{x}'),
    \\
    \frac{\delta^2 \mathcal{F}_2}{\delta \hat{C}(\mathbf{x}) \delta \hat{C}(\mathbf{x}')} 
    &= 2 \nabla' \hat{\Phi}(\mathbf{x}')\cdot\nabla' K(\mathbf{x}'-\mathbf{x}).
\end{align}
Now using the fact that 
$
\nabla' K(\mathbf{x}'-\mathbf{x}) = - \nabla K(\mathbf{x}-\mathbf{x}'),
$
we obtain an expression for the commutation of the second derivatives of the presumed free energy:
\begin{align} \label{eq:commute2deriv}
    &\frac{\delta^2\mathcal{F}_2}{\delta \hat{C} (\mathbf{x}') \delta \hat{C} (\mathbf{x})} - \frac{\delta^2\mathcal{F}_2}{\delta \hat{C} (\mathbf{x}) \hat{C} (\mathbf{x}')} = \nonumber\\
    &\quad 2 \nabla K(\mathbf{x}-\mathbf{x}') \cdot \int \mathrm{d}^d\mathbf{y} \nabla \hat{C}(\mathbf{y}) \big[K(\mathbf{x}-\mathbf{y}) +K(\mathbf{x}'-\mathbf{y})\big]. 
\end{align}
For an arbitrary particle density $\hat{C}(\mathbf{y})$, the r.h.s. cannot be identically zero for any given $\mathbf{x}\neq\mathbf{x}'$. 
We therefore conclude that since the difference between the second derivatives of the presumed free energy $\mathcal{F}_2$ does not vanish, this free energy is not well-defined. This means that, in contrast to the $\nu_1$ term in Eq.~\eqref{eq:J_DKdef}, the $\frac{\nu_2 \chi}{6 D_r}$ term cannot be derived from an underlying functional form and hence breaks the condition of detailed balance.



\section{Power counting for a generic interaction term}     \label{app_powerCounting}
    
    
Based on the engineering dimensions of the fields derived in Sec. \ref{sec_scaling} (see Eq.~\eqref{eq_fieldDimensions}), here we determine the relevance of all possible interaction terms that may be generated by the RG flow of the Langevin equation \eqref{eq:langevin} by considering their scaling behavior. 

The most general form of an interaction term can be written symbolically as
\begin{equation}    \label{generalCoupling}
    g_{lmn}~\nabla^l~(\nabla\phi)^m~\rho^n  \, , 
\end{equation}
where $\phi$ only appears together with a gradient operator to ensure the symmetry $\phi \to \phi + \mathrm{const}$, as the absolute value of the chemical field does not affect the dynamics of the particles. 
For this general coupling we assume that $m$ and $n$ are nonnegative integers (such that the resulting equation is local in space in terms of these fields) and  $m+n>0$ (in order to have at least one field involved in the coupling). 
In addition, considering the conserved dynamics of the density fluctuations, we have $l \geq 1$ to ensure that the interaction terms come as the divergence of a vector field.
Finally, the interaction terms appearing in Eq.~\eqref{eq:langevin} should be a scalar density and therefore the sum $l+m$ must be even, which also guarantees the invariance of the resulting term under spatial inversion (due to their physical meaning, $\rho$ and $\phi$ are expected to be invariant under inversion). 

The engineering dimension of the coupling $g_{lmn}$ added to the Langevin equation~\eqref{eq:langevin} can be computed by using Eqs.~\eqref{eq_fieldDimensions}. In the case of conserved noise, we obtain
\begin{align}
    [g_{lmn}]^{\rm con}_0=2+m-l-\frac{d}{2}(m+n-1) \, ,
    \label{eq_generalDimension_conserved}
\end{align}
whereas with a nonconserved noise, the engineering dimension reads as
\begin{align}
    [g_{lmn}]^{\rm non}_0=1+2m+n-l-\frac{d}{2}(m+n-1) \, .
    \label{eq_generalDimension_nonconserved}
\end{align}

The expressions in Eqs.~\eqref{eq_generalDimension_conserved}  and~\eqref{eq_generalDimension_nonconserved} allow us to identify all the terms that are marginal or relevant at the upper critical dimension. The relevant or marginal terms are the same in both of the conserved and nonconserved cases and are displayed and commented on in Table~\ref{table_relevantdc}. 
By examining the possible interaction term $g_{lmn}$, we find that in addition to the $\mu_{1,2}$ couplings included in Eq.~\eqref{eq:langevin}, which are both equally relevant and also Galilean symmetric, there exists another independent and relevant term, namely $\mu_4 \nabla \cdot (\nabla\phi)^3$. 
Using Eqs.~\eqref{eq_generalDimension_conserved}  and~\eqref{eq_generalDimension_nonconserved}, one can show that this coupling has an engineering dimension:
\begin{align}
    [\mu_4]^{\rm con}_0 = 4- d\, , \qquad \text{and} \qquad [\mu_4]^{\rm non}_0 = 6- d \,
\end{align}
in the conserved noise and nonconserved noise cases, respectively, and is therefore relevant in both cases. 
However, this coupling is \textit{not} Galilean invariant and we discuss the implications of this symmetry breaking term at the end of Sec.~\ref{sec:results}.

Note finally that the term $\mu_3 \nabla \cdot(\rho \nabla (\nabla \phi)^2)$ that has been discarded to obtain the Langevin equation~\eqref{eq:langevin} scales as $[\mu_3]^{\rm con}_0=2-d$ and $[\mu_3]^{\rm non}_0=4-d$ and is therefore irrelevant for both noises close to (and below) their corresponding upper critical dimension, $d_c^{\rm con}=4$ and $d_c^{\rm non}=6$.

\begin{table*}
\centering
        \begin{tabular}{p{0.4cm}p{0.4cm}p{0.4cm}p{3.8cm}p{1.7cm}p{9.5cm}}
    \hline\hline
         \multirow{2}{0.4cm}{$n$} & \multirow{2}{0.4cm}{$m$} & \multirow{2}{0.4cm}{$l$}& \multirow{2}{3.8cm}{\textbf{Form of the coupling}}   & \textbf{Galilean} & \multirow{2}{0.5cm}{\textbf{Comments}}   \\
         & & & & \textbf{invariant} & \\
    \hline
        \multirow{4}{0.4cm}{0} & 1 & 1 & $\nabla^2\phi $  & yes &  Equivalent to the linear term $\rho$. \\
        & 1 & 3 & $\nabla^3\nabla\phi $  & yes &  Equivalent to the diffusion term $\nabla^2\rho$ (marginal in all dimensions). \\
        & 2 & 2 & $\nabla^2 (\nabla\phi)^2 $  & yes & 
        polarity-induced chemotactic term. \\
        & 3 & 1 & $\nabla (\nabla\phi)^3$ 
        & no & 
        Single-particle self-propulsion/nematic order\\
    \hline
        \multirow{2}{0.4cm}{1} & 0 & 2 & $\nabla^2\rho $  & yes & Diffusion term (marginal in all dimensions). \\
         & 1 & 1 & $\nabla(\rho \nabla\phi) $   & yes & KS chemotactic term. \\
    \hline\hline
        \end{tabular}
        \caption{Marginal and relevant couplings at the upper critical dimension ($d^{\rm con}_c=4$ and $d^{\rm non}_c=6$). We consider coupling of the form $\nabla^l~ (\nabla\phi)^m~\rho^n$ 
        with nonnegative $m$ and $n$ and with $m+n>0$, $l\geq 1$, and $l+m$ even.}
        \label{table_relevantdc}
\end{table*}


\section{Details of the RG calculations} \label{app_rgprocedure}

In this appendix, the details of the RG calculations of the propagator, noise, and vertex are provided. The Ward identity, which is the result of the Galilean symmetry of the Langevin equation~\eqref{eq:langevin}, is discussed at the end. In this appendix we focus on the case of a conserved noise ($\mathcal{D}_0=0$), as the same procedure applies to nonconserved noise with the resulting flow equations reported in Appendix \ref{app_rgnon}.


\subsection{Renormalization of the propagator}

We recall the diagrammatic representation of the propagator renormalization at one-loop:
\begin{center}
\tikzset{
    midarrow/.style={
        decorate,decoration={
            markings,
            mark=at position 1 with{\arrow{#1};}
        }
    }
}

\begin{tikzpicture}
[baseline={([yshift=-.5ex]current bounding box.center)},vertex/.style={anchor=base,
    circle,fill=black!25,minimum size=18pt,inner sep=2pt}]
\draw [double distance=1.5pt] (-.1,-1.5) -- (.7,-1.5);    
\draw [double distance=1.5pt, -{Latex[length=3.25mm,reversed]}] (-1,-1.5) -- (0,-1.5);
\node at (-.2,-2.5) {$G$};
\node at (1,-1.5) {$=$};
\node at (1,-2.5) {$=$};
\draw [thick, -{Latex[length=3.25mm,reversed]}] (1.5,-1.5) -- (2.5,-1.5);
\draw [thick] (2.4,-1.5) -- (3.2,-1.5);
\node at (2.4,-2.5) {$G_0$};
\node at (3.5,-1.5) {$+$};
\node at (3.5,-2.5) {$+$};
\draw [thick] (4.5,-1.5) -- (4.9,-1.5);
\draw [thick, -{Latex[length=3.25mm,reversed]}] (4.,-1.5) -- (4.6,-1.5);
\node at (4.5,-2.5) {$G_0$};
\draw[thick] (5.4,-1.5) circle (0.5cm);
\draw [thick] (6.4,-1.5) -- (6.9,-1.5);
\draw [thick, -{Latex[length=3.25mm,reversed]}] (5.9,-1.5) -- (6.6,-1.5);
\node at (6.4,-2.5) {$G_0$};
\filldraw[fill=white] (5.4,-1) circle (2pt);
\node at (5.5,-2.5) {$\Sigma_1$};
\draw [thick, -{Latex[length=2.5mm,reversed]}] (4.9,-1.5) arc (180:120:0.5cm);
\draw [thick, -{Latex[length=2.5mm,reversed]}] (5.9,-1.5) arc (0:60:0.5cm);
\draw [thick, -{Latex[length=2.5mm,reversed]}] (4.9,-1.5) arc (180:290:0.5cm);
\end{tikzpicture}
\end{center}
The loop integral $\Sigma_1$ shown above is an integral over the ``fast'' modes that reads:
\begin{align}
\begin{split}
\Sigma_1(\hat k)= \frac{8}{2!} \int_{\hat q}^> & \mathcal{N}_0(\hat k/2 + \hat q)
\Gamma_0(\mathbf k, \mathbf k/2+\mathbf q) \times \\
&\Gamma_0(\mathbf k/2-\mathbf q,\mathbf k) G_0(\hat k/2 - \hat q)
\, , 
\end{split}
\end{align}
where $\mathcal{N}_0$ is defined in Eq.~\eqref{eq:barecorr} and we have defined 
\begin{align}
\int_{\hat q}^>\equiv\int_{-\infty}^\infty \frac{\mathrm{d} \omega}{2\pi} \int_{\Lambda/b\leq |\mathbf{q}| \leq \Lambda} \frac{\mathrm{d}^d \mathbf{q}}{(2\pi)^d} \,. 
\end{align}

To compute the renormalization of $\sigma$ and $D$, it is more convenient to consider the renormalization of the inverse propagator $G^{-1}$ which is given by the Dyson expansion  \cite{tauber2014}: $G^{-1}(\hat k) = G_0^{-1}(\hat k)-\Sigma_1(\hat k)$. The renormalized coupling constants  $\sigma_{\rm R}$ and $D_{\rm R}$ of the propagator are then computed as:
\begin{align}
\sigma_{\rm R} & = G^{-1}(\hat{k}) |_{\hat{k}=0}\, , \\
D_{\rm R} &= \partial_{k^2} G^{-1}(\hat k) \bigg\rvert_{\hat k= \hat 0} \, .  
\end{align}
The explicit computation of the loop integral $\Sigma_1$ is done by first computing the integral over the frequencies using residues. The $d$-dimensional integral over the internal momentum $\mathbf{q}$ is then reduced to a one dimensional integral over its norm $|\mathbf{q}|=q$ 
\charlie{by making use of the angular symmetry around $\bm{k}$.}
Finally, the integration over the norm itself is performed in the limit where $b\equiv \mathrm{e}^{\delta\ell}$ is infinitesimally close to 1 and, thus, $\int_q^>f(q) = f(\Lambda)\Lambda \delta\ell +\mathcal{O}(\delta\ell^2)$. In the conserved case ($\mathcal{D}_0=0$), this gives at one-loop:
\begin{align}
    \sigma_{\rm R} & = \sigma \, , \\
    D_{\rm R} &= D - \frac{K_d \delta\ell \Lambda^{d-4} \mathcal{D}_2}{D^2} \left( a_{11} \mu_1^2 + a_{12} \mu_1 \mu_2 + a_{22} \mu_2^2 \right) \, ,
\end{align}
where $K_d=2/[(4\pi)^{d/2}\Gamma(d/2)]$ and the coefficients $a_{11} = 3/4 - 3/(2 d)$,  $a_{12} = 2+3/d-6/(d+2)$, $a_{22} = 1-4/d$ are the same as those introduced in the main text below Eq.~\eqref{eq:cnoiseflows}. 


Performing the integration over the ``fast'' modes gives the renormalized coupling constants with $\Lambda/b$ as the momentum cutoff. To restore the original cutoff $\Lambda$, we rescale space, time and the fluctuation field according to Eq.~\eqref{eq_scaling}.
This rescaling completes the RG calculation, with the new coupling constants expressed in terms of the old ones. In the limit where the change of scale is infinitesimal (that is $b=\mathrm{e}^{\delta\ell}$ with $\delta \ell \ll 1$), the change of the coupling constants $\sigma$ and $D$ under the RG step can be cast into a set of coupled differential equations, which are the RG flow equations displayed in the main text, Eqs.~\eqref{eq:cnoiseflowsigma} and~\eqref{eq:cnoiseflowD}.


\subsection{Renormalization of the noise}
        
The renormalization of the dynamic correlation function $\mathcal{N}$ is performed using the diagrammatic representation shown in Fig.~\ref{fig:diagrams}. Calling $\mathcal{N}_1$ the one-loop contribution, we have:%
\begin{equation}
\begin{split}
    \mathcal{N}_1 = \frac{4}{2!} \int_{\hat q}^> &\mathcal{N}_0(\hat k/2 + \hat q) \mathcal{N}_0(\hat k/2 - \hat q)  \\
    &\times\Gamma_0(\mathbf k, \mathbf k/2+\mathbf q) \Gamma_0(-\mathbf k, -\mathbf k/2-\mathbf q) \, ,
\end{split}
\end{equation}
from which we can extract the renormalized nonconserved and conserved noise terms $\mathcal{D}_{0_{\rm R}}$ and $\mathcal{D}_{2_{\rm R}}$ following the same procedure as for the propagator. 
%

In particular, one can check that the lowest term in the series expansion in $k$ of $\mathcal{N}_1$ goes as $k^4$, as discussed in the main text. Indeed, the series read:
\begin{align}
\begin{split}
    &\mathcal{N}_1 = k^4 \frac{(\mathcal{D}_0 + \Lambda^2\mathcal{D}_2)^2 K_d \Lambda^{d-10}\delta \ell }{D^3} \times\\
    &\!\left[\! \left( \frac{1}{4} \!+\! \frac{1}{2d} - \frac{3}{2(2+d)}\!\right)\! \mu_1^2 +\!\left(\!1 - \frac{2}{d}\right)\!\mu_1\mu_2 + \mu_2^2  \right] \!+\! \mathcal{O}(k^6),
\end{split} \label{eq_noiseExpansion}
\end{align}
and there is no contribution in $k^0$ or $k^2$ that could renormalize the nonconserved or conserved noise, respectively. The second part of the RG step (rescaling) can then be performed as described in the case of the propagator, and we obtain Eq.~\eqref{eq:cnoiseflowD2} in the conserved noise case.


\subsection{Renormalization of the vertex}

%
%
The diagrammatic representation of the vertex renormalization is shown in Fig.~\ref{fig:diagrams}. In addition to the bare diagram, there are three diagrams that contributes at one-loop, whose contributions are denoted from left to right by  $\Gamma_1^{(a)}$, $\Gamma_1^{(b)}$ and $\Gamma_1^{(c)}$ and read:
\begin{align}
\begin{split}
    &\Gamma_1^{(a)}(\mathbf k, \mathbf k/2+\mathbf p) = 4 \int_{\hat q}^>  \mathcal{N}_0(\hat k/2 + \hat q)
    \Gamma_0(\mathbf k, \mathbf k/2 + \mathbf q) \times \\
    &\hskip2cm \Gamma_0(\mathbf p - \mathbf q, \mathbf k/2 + \mathbf p) \Gamma_0(\mathbf k/2 - \mathbf q, \mathbf p - \mathbf q) \times \\
    &\hskip2cm  G_0(\hat p - \hat q) G_0(\hat k/2 - \hat q) \, , 
\end{split}\\
\begin{split}
    &\Gamma_1^{(b)}(\mathbf k, \mathbf k/2+\mathbf p) = 4 \int_{\hat q}^>  \mathcal{N}_0(\hat k/2 - \hat q) \Gamma_0(\mathbf k, \mathbf k/2 + \mathbf q) \times \\
    &\hskip2cm \Gamma_0(\mathbf k/2 + \mathbf q, \mathbf k/2 + \mathbf p) \Gamma_0(\mathbf q - \mathbf p, \mathbf k/2 - \mathbf p) \times \\
    &\hskip2cm G_0(\hat k/2 + \hat q) G_0(\hat q - \hat p) \, , 
\end{split}\\
\begin{split}
    &\Gamma_1^{(c)}(\mathbf k, \mathbf k/2+\mathbf p) = 4 \int_{\hat q}^> \mathcal{N}_0(\hat p - \hat q)
    \Gamma_0(\mathbf k, \mathbf k/2 + \mathbf q) \times \\
    &\hskip2cm \Gamma_0(\mathbf k/2 + \mathbf q, \mathbf k/2 + \mathbf p) \Gamma_0(\mathbf k/2 - \mathbf q, \mathbf k/2 - \mathbf p) \times \\
    &\hskip2cm G_0(\hat k/2 + \hat q) G_0(\hat k/2 - \hat q) \, . 
\end{split}
\end{align}
In order to compute the renormalization of the chemotactic terms $\mu_{1,2}$, the dependency of $\Gamma_1 =\Gamma_1^{(a)}+\Gamma_1^{(b)}+\Gamma_1^{(c)}$ on the external momenta $\mathbf k$ and $\mathbf p$ has to be kept. We first compute the frequency integral appearing in $\Gamma_1$ using residues. Then, we focus on the ultraviolet divergence (when $\Lambda\to\infty$) of $\Gamma_1$ and compute the residue of the pole in $1/q^{d_c}$, which gives rise to the renormalization of the coupling constants $\mu_{1,2}$ at the critical point. At one-loop, this residue vanishes, which yields $\mu_{1,\rm R}=\mu_1$ and $\mu_{2,\rm R}=\mu_2$. The second part of the RG step (rescaling) can then be performed as described in the case of the propagator, and we obtain Eq.~\eqref{eq:cnoiseflowmu12} in the conserved noise case.


\subsection{Galilean symmetry and Ward identity}    \label{app:ward}


The Galilean symmetry \eqref{eq:galileantransform} discussed in the main text implies that the term $\mu_1-2\mu_2$ remains constant along the RG flow, and yields the exponent identity~\eqref{eq:exponidentity}. The invariance of the term $\mu_1-2\mu_2$ along the RG flow can be made more formal by looking at the Ward identity associated to this symmetry~\cite{tauber2014,frey1994}. The Ward identity expresses a relation between the three-point vertex function $\Gamma$ and the two-point vertex function (or inverse propagator) $G^{-1}$ that reads: 
%
%
\begin{align}
    i\,\left(\mu_1-2\mu_2\right) \, \mathbf{q} \, \partial_{\omega} G^{-1}(\hat{q}) = \partial_{\mathbf{k}} \, \Gamma (\hat{k} \,;\hat{q})\big\rvert_{\hat{k}=0}.
    \label{eq:ward}
\end{align}
We thus conclude, similarly to the KPZ case~\cite{frey1994,canet2011a}, that $\mu_1-2\mu_2$ is not renormalized and remains equal to its bare value.


\section{RG flow equations for the nonconserved noise}  \label{app_rgnon}
    
Following the procedure described in Appendix~\ref{app_rgprocedure}, one can also treat the nonconserved noise case ($\mathcal{D}_0\neq0$). As discussed in the main text, the conserved part of the noise is irrelevant in this case and will be discarded from the analysis. The final RG equations for the nonconserved noise case read as
\begin{subequations}
\label{eq:nnoiseflows}
\begin{align}
\partial_{\ell} \sigma &= \left[d+2\chi \right] \sigma ,\\ 
\partial_{\ell} \mu_{1,2} &=  \left[ z+\chi \right]\mu_{1,2}, \label{eq:nnoiseflowmu12}\\[1.5mm]
\partial_{\ell} D \,\, &= \Big[ z-2 - \big( b_{11} U_1^2 
     + b_{12} U_1 U_2 + b_{22} U_2^2\big)\Big] D,\label{eq:nnoiseflowD}
 \\[1.5mm]
\partial_\ell \mathcal{D}_0 &=  \left[-d+z-2\chi\right] \mathcal{D}_0, \label{eq:nnoiseflowD0}
\end{align}
\end{subequations}
where we have defined $U_{1,2}^2 = \mu_{1,2}^2 \, \mathcal{D}_0 \, K_d \Lambda^{d-6}  / D^3$ 
and the coefficients  
$b_{11} = 3/4 - 1/d - 3/[d(d+2)]$, 
$b_{12} = 2 + 6/d - 9/(d+2)$, and 
$b_{22} = 1-6/d$.


\section{Analysis of the renormalization group flows in various dimensions}  \label{app_rgflowsdim}

In this section, we look into the structure of the renormalization group flows in various spatial dimensions $d$. We remind the reader that despite the scaling exponents that are obtained exactly, the RG flow equations and the corresponding analysis are only valid to one-loop order and a higher order calculation will be required to form a more conclusive picture of different phases of the system in the parameter space.


\subsection{Structure of the fixed-point solutions}

The RG flows for the effective couplings $U_{1,2}$ in the presence of conserved and nonconserved noise are given by Eq.~\eqref{eq:cnoiseflowU1U2} and Eq.~\eqref{eq:nnoiseflowU1U2}, respectively. In both cases, the r.h.s is the same for both $\partial_\ell U_1$ and $\partial_\ell U_2$, indicating that the flows occur along the rays with a fixed ratio of $U_2/U_1$. The fixed points are obtained by setting $\partial_\ell U_{1,2}=0$ which, besides the trivial Gaussian fixed point $U_1=U_2=0$, results in a quadratic equation 
$A U_1^2 + B U_{1} U_{2} + C U_2^2 + E = 0$ 
where the coefficients $A, B, C$, and $E$ are defined below Eqs.~\eqref{eq:cnoiseflowU1U2} and \eqref{eq:nnoiseflowU1U2} in each case. This quadratic equation defines conic sections in the $U_1$-$U_2$ plane whose shape can be determined based on the sign of its discriminant $\Delta$ defined as
\begin{align}
    \Delta = B^2 - 4 A C.
\end{align}

The fixed points in various dimensions and their shape are shown in Fig.~\ref{fig:rgflowvardim}.


\subsection{Linear stability analysis of the fixed-point curves}

To analyze the stability of the lines of (nontrivial) fixed-point, we consider a small displacement from a fixed-point $(U_1^*,U_2^*)$ to the neighbouring point $(U_1^*+\delta U_1, U_2^*+\delta U_2)$. Since the flows are along the rays passing through the origin, we assume the displacement is also along the ray passing through the initial point, i.e. $\delta U_1/U_1^* = \delta U_2/U_2^*$, so that if the fixed point is attractive the RG flow will return to the same state. Expanding the flow equations~\eqref{eq:cnoiseflowU1U2} and \eqref{eq:nnoiseflowU1U2} to the leading order in $\delta U_1$ and $\delta U_2$, we get in both cases
\begin{align}
    \partial_\ell \left(\delta U_{1,2}\right) \big\rvert_{(U_1^*,U_2^*)} = -2 E \left(\delta U_{1,2}\right),
\end{align}
where $E= (d_c - d)/2$ in both cases of conserved and nonconserved noise with
$d_c = d_c^{\mathrm{con}}=4$
and
$d_c = d_c^{\mathrm{non}}=6$,
respectively. It is therefore clear that for $d<d_c$, the flows are attractive and the nontrivial fixed points are stable, whereas for $d>d_c$ the flows are repulsive and the nontrivial fixed points become unstable. Exactly at the upper critical dimension $d_c$ the nontrivial fixed points form straight lines and they become neutral (in the sense of stability that is considered here).

\color{black}


\section{Details of moment expansion for more general chemotactic  mechanisms} \label{app:moment_long}


In this appendix we extend the calculation that was presented in Appendix~\ref{app:moment} to take into account the self-propulsion and nematic alignment of the particles.
To this end, consider the more general case of Eqs.~\eqref{eq:microTLangevin} and \eqref{eq:microRLangevin} as
\begin{align}   
    \frac{\mathrm{d}}{\mathrm{d}t}\mathbf{r}_a(t) &= \mathbf{v} \left(\mathbf{r}_a,\mathbf{n}_a\right) + \bm{\xi}_a(t), 
    \\
    \frac{\mathrm{d}}{\mathrm{d}t}\mathbf{n}_a(t) &= \bm{\omega} \left(\mathbf{r}_a,\mathbf{n}_a\right) + \bm{\gamma}_a(t)\times\mathbf{n}_a.
\end{align}
One can formulate expressions for $\mathbf{v}$ and $\bm{\omega}$ based on a general gradient expansion  \cite{saha2014,golestanian2012}, which read 
%
\begin{align} \label{eq:v-for-moment_long}
    \mathbf{v} =  v_0\mathbf{n} + \nu_1 \nabla\Phi + \nu_2 \mathbf{n}\cdot\nabla \nabla\Phi + \nu_3 \mathbf{n}\mathbf{n}\cdot\nabla\Phi,
\end{align}
and Eq.~\eqref{eq:microRLangevin} for the polarity. 
To obtain these expressions, one has to assume that the particles have a linear measurement mechanism to sense the chemical gradient.
Note that in comparison with Eq.~\eqref{eq:microTLangevin}, we have added the $v_0$ term to the translational Langevin equation to include the case of self-propelling particles, 
and the $\nu_3$ term in order to include the cases where a nonvanishing local nematic order exists, which could arise, for instance, from geometric asymmetries of the particles~\cite{saha2014}. 
Note that we have assumed for simplicity that the 
polarity vector $\mathbf{n}$ also defines the self-propulsion direction, although these two directions need not be parallel in general.
As before, the noise terms $\bm{\xi}_a$ and $\bm{\gamma}_a$ are Gaussian white noises acting on the $a$th particle, characterized by Eq.~\eqref{eq:microTnoise} while we now modify the 
polarity part as
\begin{align}   \label{eq:microRnoise_long}
\langle \gamma_{al}(t) \gamma_{bm}(t') \rangle = 2 (D_r-g v_0 \mathbf{n}_a \cdot \nabla \Phi)\delta_{ab}\delta_{lm}\delta(t-t'), 
\end{align}
which generalizes Eq.~\eqref{eq:microRnoise} to the case of run-and-tumble particles by modulating the tumble rate due to the chemical gradient  
(see Ref. \cite{schnitzer1993}). 

The Fokker--Planck equation in this case reads as
\begin{equation} \label{eq:FP_appendix_long}
\begin{split}
\partial_t\mathcal{P}=&- \nabla\cdot\left[ \mathbf{v}\mathcal{P} - D\nabla\mathcal{P}\right]\\
    &- \bm{\mathcal{R}}\cdot\left[\bm{\omega}\mathcal{P}-\bm{\mathcal{R}} \big((D_r-g v_0 \mathbf{n}\cdot \nabla \Phi)\mathcal{P}\big)\right], 
\end{split}
\end{equation}
The 
noise corresponding to the polarity has been implemented here using the Ito convention, which is the appropriate choice given the discrete nature of the run-and-tumble process. 


The hierarchy of equations governing the moments of the distribution $\mathcal{P}$ with respect to $\mathbf{n}$ can now be constructed.
Performing the integration of Eq.~\eqref{eq:FP_appendix_long} with respect to the polarity $\mathbf{n}$ leads to the continuity equation \eqref{eq:continuity_meanfield}
where the particle current is now given by 
\begin{equation} \label{eq:J_meanfielddef_appendix_long}
\begin{split}
\mathbf{J} (\mathbf{x},t) =
    -D\nabla C &+ v_0 \bm{p} +  \left(\nu_1 \!+\! \frac{\nu_3}{3}\right) C \nabla\Phi \\ 
    &  + \nu_2 \bm{p}\!\cdot\!\nabla\nabla\Phi + \nu_3 \bm{Q}\!\cdot\!\nabla\Phi ,
\end{split}
\end{equation}
where the density and polarization fields are defined in Eqs.~\eqref{eq:C_meanfielddef} and \eqref{eq:p_meanfielddef}, and the nematic order field is given by 
\begin{align}   \label{eq:Q_def}
\bm{Q}(\mathbf{x},t) = \int_{\mathbf{n}} \left[\mathbf{n}\mathbf{n}-\frac{1}{3} \, \mathbf{I}\right]\mathcal{P}(\mathbf{x},\mathbf{n};t).
\end{align}
The dynamics of $\bm{p}(\mathbf{x},t)$ is given by 
\begin{equation} \label{eq:polarizationFP_appendix_long}
\begin{split}
    &\partial_t  p_i + \partial_l \bigg[ 
    -D\partial_l p_i + v_0\left(Q_{il} + \frac{1}{3}  C \delta_{il}\right)\\
    &\quad +\nu_1  p_i  \partial_l\Phi 
    + \nu_2 \left( Q_{ik} + \frac{1}{3} C \delta_{ik}\right)  \partial_k \partial_l \Phi\\
    &\quad + \nu_3\left( Q^{(3)}_{ilk} \partial_k \Phi + \frac{1}{5} \left( p_i \partial_l\Phi + p_l \partial_i\Phi + \delta_{il}p_k\partial_k\Phi\right)\right)
    \bigg] \\
    & \quad + 2 D_r p_i - \frac{2}{3} \left( \chi + g v_0\right) C \partial_i\Phi
    +\left(\chi - 2 g v_0 \right) Q_{il}\partial_l \Phi = 0, 
\end{split}
\end{equation}
\onecolumngrid
where we have used the definition
\begin{align}   \label{eq:Q3_def}
Q^{(3)}_{ilk} = \int_{\mathbf{n}} \mathcal{P}(\mathbf{x},\mathbf{n};t) &\bigg[n_i n_l n_k  \!-\! \frac{1}{5} \left(n_i\delta_{lk}+n_l\delta_{ik}+n_k\delta{il}\right)\bigg].
\end{align}
By continuing this procedure, the equation for the nematic order parameter field is obtained as
\begin{equation}
\begin{split}  \label{eq:nematicFP_appendix}
    & \partial_t Q_{ij} + \frac{1}{3}\partial_t C \delta_{ij}+6  D_r  Q_{ij}  -\frac{1}{5} \left(\chi+2gv_0\right) \bigg(3 \left(p_i\partial_j \Phi + p_j\partial_i \Phi\right)-2\delta_{ij} p_l \partial_l\Phi \bigg)+ 2\left(\chi-3gv_0\right) Q_{ijl}^{(3)} \partial_l \Phi \\
    &+ \partial_l \bigg[ 
    v_0 \left( Q_{ijl}^{(3)} + \frac{1}{5}\left(p_i\delta_{jl}+p_j\delta_{il}+p_l\delta_{ij}\right) \right)+\nu_1 \partial_l\Phi\left(Q_{ij} + \frac{1}{3}C \delta_{ij}\right) + \nu_2 \, \partial_k\partial_l\Phi \left(Q^{(3)}_{ijl} + \frac{1}{5}\left(p_i\delta_{jk}+p_j\delta_{ik}+p_k\delta_{ij}\right)\right) \\
    &+ \nu_3 \, \partial_k \Phi \left(Q_{ijkl}^{(4)} + \frac{1}{7} \left(Q_{ij}\delta_{kl} + Q_{ik}\delta_{lj}+Q_{il}\delta_{jk} +Q_{jk}\delta_{il}+Q_{jl}\delta_{ik}+Q_{kl}\delta_{ij}\right) + \frac{C}{15}  \left(\delta_{ij}\delta_{kl}+\delta_{ik}\delta_{jl}+\delta_{il}\delta_{jk}\right)\right)\\
    &\quad -D\partial_l\left(Q_{ij}+ \frac{1}{3}C \delta_{ij}\right)\bigg] = 0, 
\end{split}
\end{equation}
where we have defined
\begin{align}   \label{eq:Q4_def}
    Q^{(4)}_{ijkl} = \int_\mathbf{n} &\mathcal{P}(\mathbf{x},\mathbf{n};t) \bigg[n_i n_j n_k n_l - \frac{1}{7} \left( n_i n_j\delta_{kl} + n_i n_k\delta_{lj} + n_i n_l\delta_{jk} + n_j n_k\delta_{il} + n_j n_l\delta_{ik} + n_k n_l\delta_{ij} \right) \nonumber \\
    &+\frac{1}{35}\left(\delta_{ij}\delta_{kl}+\delta_{ik}\delta_{jl}+\delta_{il}\delta_{jk}\right)\bigg]. 
\end{align}
This procedure will generate a hierarchy of equations involving higher order moments of the distribution function. 
As in the case of the celebrated  Bogoliubov-Born-Green-Kirkwood-Yvon (BBGKY) hierarchy in liquid state theory \cite{born1946,kardar2007}, the hierarchy can be truncated by using a closure scheme. 
Here, we close the hierarchy by assuming that $Q^{(3)}$, $Q^{(4)}$, and all higher order moments vanish. 
Since we are interested in the macroscopic properties of the system, we employ a hydrodynamic approximation and focus on time scales much longer than $D_r^{-1}$ and length scales much larger than $\sqrt{D/D_r}$. 
This allows us to further simplify Eqs.~\eqref{eq:polarizationFP_appendix_long} and \eqref{eq:nematicFP_appendix} to obtain expressions for the polarization field and the nematic tensor~\cite{golestanian2019}.  
In this limit, we obtain
\begin{align}
    p_i = &-\left(\frac{v_0}{6 D_r}\right) \partial_i C 
    + \left(\frac{\chi+gv_0}{3 D_r}\right) C \, \partial_i\Phi \nonumber\\
    &+ \left(\frac{v_0\left(\nu_1 + \nu_3/5\right)}{12 D_r^2} \right) \partial_i C \, \partial_l^2\Phi 
    -  \left(\frac{\left(\chi+gv_0\right)\left(\nu_1+\nu_3/5\right)}{6 D_r^2} \right) C \, \partial_i\Phi \, \partial_l^2\Phi 
    -\left(\frac{\nu_2}{6 D_r}\right) 
    \bigg[\partial_lC \, \partial_i\partial_l\Phi + C \, \partial_i\partial_l^2\Phi\bigg]\nonumber\\
    & + \left(\frac{D}{2 D_r} \right) \partial_l^2 p_i 
    - \left(\frac{\nu_1}{ 2 D_r} \right) \partial_l p_i \, \partial_l\Phi
    - \left(\frac{\nu_3}{10 D_r}\right) \bigg[\partial_l p_i\, \partial_l\Phi + \partial_l p_l \, \partial_i\Phi + 2 p_l \, \partial_i\partial_l\Phi + \partial_i p_l \, \partial_l\Phi\bigg] \nonumber\\
    & -\left(\frac{v_0}{2 D_r} \right) \partial_l Q_{il} 
    - \left( \frac{\chi-2gv_0}{2 D_r}\right) Q_{il} \, \partial_l\Phi + \mathcal{O}(\nabla^5),
\end{align}
and
\begin{align}
    Q_{ij} = &- \left(\frac{\nu_3}{90 D_r}\right) \left[ \partial_i\Phi \, \partial_jC + 2 C \, \partial_i\partial_j\Phi + \partial_iC \, \partial_j\Phi \right] + 
    \delta_{ij} \left( \frac{\nu_3}{135 D_r}\right) \left[ \partial_l C \, \partial_l\Phi + C \partial_l^2 \Phi \right]  \nonumber \\
    & -\left(\frac{v_0}{30 D_r}\right) \left[\partial_i p_j + \partial_j p_i \right] 
    + \delta_{ij} \left(\frac{v_0}{45 D_r} \right) \partial_l p_l
    + \left(\frac{\chi +2 gv_0}{10 D_r}\right) \left[p_i \, \partial_j\Phi + p_j \, \partial_i\Phi\right] 
    - \delta_{ij} \left( \frac{\chi+2gv_0}{15 D_r} \right) p_l\partial_l\Phi
    +\mathcal{O}(\nabla^4).
\end{align}
We can now use the above expressions to solve for $p_i$ and $Q_{ij}$ in terms of the scalar fields only. 
This calculation yields
\begin{equation}    \label{eq:Q_order2_appendix_long}
    \begin{split}
        Q_{ij} = \, &\left(\frac{v_0^2}{90 D_r^2} \right) \partial_i\partial_j C
        -\left(\frac{v_0 \left(5\chi+8gv_0\right)}{180 D_r^2} + \frac{\nu_3}{90 D_r} \right) \left[\partial_iC \, \partial_j\Phi + \partial_jC \, \partial_i\Phi\right] 
        -\left(\frac{v_0\left(\chi+gv_0\right)}{45D_r^2} + \frac{\nu_3}{45 D_r} \right) C \, \partial_i\partial_j\Phi  \\
        & + \left(\frac{\left(\chi+gv_0\right)\left(\chi+2gv_0\right)}{15 D_r^2} \right) C \, \partial_i\Phi \, \partial_j\Phi
        +\delta_{ij} \left(\frac{v_0 \left(5\chi+8gv_0\right)}{270 D_r^2} + \frac{\nu_3}{135 D_r} \right) \partial_lC \, \partial_l \Phi \\
        &-\delta_{ij} \left( \frac{v_0^2}{270 D_r^2} \right) \partial_l^2C 
        + \delta_{ij}\left(\frac{v_0\left(\chi+gv_0\right)}{135 D_r^2} + \frac{\nu_3}{135 D_r}\right)C \, \partial_l^2\Phi 
        -\delta_{ij} \, \left(\frac{\left(\chi+gv_0\right)\left(\chi+2gv_0\right)}{45 D_r^2}\right) C \, \left(\partial_l\Phi\right)^2 
        +\mathcal{O}(\nabla^4),
    \end{split}
\end{equation}
and
\begin{equation}  \label{eq:p_order2_appendix_long}
    \begin{split}
    p_i = 
    & - \left(\frac{v_0}{6 D_r} \right) \partial_iC 
    + \left(\frac{\chi+gv_0}{3 D_r} \right)C \, \partial_i\Phi 
    + \left(\frac{D(\chi+gv_0)}{6 D_r^2} +\frac{v_0^2 (\chi+gv_0)}{135 D_r^3}-\frac{\nu_2}{6 D_r} + \frac{v_0\nu_3}{135 D_r^2} \right)  C\, \partial_i\partial_l^2\Phi \\
    &- \left( \frac{\nu_1(\chi+gv_0)}{6 D_r^2} + \frac{\nu_3(19\chi+16gv_0)}{270 D_r^2}
    +\frac{v_0 (\chi+gv_0)(5\chi+8gv_0)}{135 D_r^3} \right) C \, \partial_i\Phi \, \partial_l^2\Phi \\
    & - \left( \frac{(\chi+gv_0)\nu_1}{6 D_r^2} + \frac{(11\chi+14 gv_0)\nu_3}{90 D_r^2} + \frac{2gv_0^2(\chi+gv_0)}{45 D_r^3} \right)C \,\partial_l\Phi \, \partial_l\partial_i\Phi \\
    &-\left( \frac{(\chi-2gv_0)(\chi+gv_0)(\chi+2gv_0)}{45 D_r^3} \right) C\,\partial_i\Phi (\partial_l\Phi)^2 
    + \left( \frac{v_0\nu_1}{12 D_r^2} + \frac{v_0 \nu_3}{54 D_r^2} + \frac{v_0^2(11\chi+20gv_0)}{1080 D_r^3} \right) \partial_iC \, \partial_l^2\Phi \\
    & + \left(\frac{D(\chi+gv_0)}{3 D_r^2} + \frac{v_0^2(17\chi+20gv_0)}{1080 D_r^3}-\frac{\nu_2}{6D_r} + \frac{5 v_0 \nu_3}{108 D_r^2} \right) \partial_lC \,\partial_i\partial_l\Phi \\
    & + \left( \frac{v_0(9\chi^2 + 10\chi g v_0 - 8g^2v_0^2)}{360 D_r^3} - \frac{\nu_3(5\chi+8gv_0)}{180 D_r^2} \right) \, \partial_iC (\partial_l\Phi)^2 \\
    &- \left(\frac{v_0(31\chi^2 + 110\chi g v_0 + 88 g^2 v_0^2)}{1080 D_r^3} + \frac{(\chi+gv_0)\nu_1}{6 D_r^2} + \frac{\nu_3(35\chi+38 gv_0)}{540 D_r^2} \right) \partial_lC \, \partial_l\Phi \, \partial_i\Phi \\
    & + \left( \frac{D(\chi+gv_0)}{6 D_r^2} + \frac{v_0^2(17\chi+20gv_0)}{1080 D_r^3} + \frac{v_0 \nu_3}{45 D_r^2} \right) \partial_l^2C \, \partial_i\Phi
    + \left( -\frac{v_0^2(\chi - 20 g v_0)}{1080 D_r^3} + \frac{v_0\nu_1}{12 D_r^2} + \frac{19 v_0 \nu_3}{540 D_r^2} \right) \partial_l\partial_iC \, \partial_l\Phi \\
    &  - \left( \frac{v_0 D}{12 D_r^2} + \frac{v_0^3}{270 D_r^3}\right) \partial_i\partial_l^2C+\mathcal{O}(\nabla^5).
    \end{split}
\end{equation}
Finally, we can derive the following expression for current in terms of the scalar fields only
\begin{equation}  \label{eq:J_final_appendix_long}
    \begin{split}
    \mathcal{J}_i = 
    & - \left(D + \frac{v_0^2}{6 D_r} \right) \partial_iC 
    + \left(\nu_1 + \frac{\nu_3}{3} + \frac{v_0(\chi+gv_0)}{3 D_r} \right)
    C \, \partial_i\Phi \\
    &+ \left(\frac{v_0 D(\chi+gv_0)}{6 D_r^2} +\frac{v_0^3 (\chi+gv_0)}{135 D_r^3}-\frac{v_0\nu_2}{6 D_r} + \frac{v_0^2\nu_3}{135 D_r^2} \right)  
    C\, \partial_i\partial_l^2\Phi \\
    &- \left(-\frac{\nu_3^2}{135 D_r} + \frac{v_0\nu_1(\chi+gv_0)}{6 D_r^2} + \frac{v_0\nu_3(17\chi+14gv_0)}{270 D_r^2}
    +\frac{v_0^2 (\chi+gv_0)(5\chi+8gv_0)}{135 D_r^3} \right) 
    C \, \partial_i\Phi \, \partial_l^2\Phi \\
    & - \left( -\frac{\nu_2(\chi+gv_0)}{3D_r} + \frac{\nu_3^2}{45 D_r} +\frac{v_0\nu_1(\chi+gv_0)}{6 D_r^2} + \frac{v_0\nu_3(13\chi+16 gv_0)}{90 D_r^2} + \frac{2gv_0^3(\chi+gv_0)}{45 D_r^3} \right)
    C \,\partial_l\Phi \, \partial_l\partial_i\Phi \\
    &+\left( \frac{2\nu_3(\chi+gv_0)(\chi+2gv_0)}{45 D_r^2} -\frac{v_0(\chi-2gv_0)(\chi+gv_0)(\chi+2gv_0)}{45 D_r^3} \right) C\,\partial_i\Phi (\partial_l\Phi)^2  \\
    &+ \left( \frac{v_0^2\nu_1}{12 D_r^2} + \frac{v_0^2 \nu_3}{54 D_r^2} + \frac{v_0^3(11\chi+20gv_0)}{1080 D_r^3} \right) 
     \partial_iC \, \partial_l^2\Phi \\
     & + \left(\frac{v_0 D(\chi+gv_0)}{3 D_r^2} + \frac{v_0^3(17\chi+20gv_0)}{1080 D_r^3}-\frac{v_0 \nu_2}{3D_r} + \frac{5 v_0^2 \nu_3}{108 D_r^2} \right) \partial_lC \,\partial_i\partial_l\Phi \\
     & + \left( \frac{v_0^2 (9\chi^2 + 10\chi g v_0 - 8g^2v_0^2)}{360 D_r^3} - \frac{v_0\nu_3(5\chi+8gv_0)}{90 D_r^2} -\frac{\nu_3^2}{90 D_r}\right)
     \partial_iC (\partial_l\Phi)^2 \\
     & + \left( \frac{v_0 D(\chi+gv_0)}{6 D_r^2} + \frac{v_0^3(17\chi+20gv_0)}{1080 D_r^3} + \frac{v_0^2 \nu_3}{54 D_r^2} \right) 
     \partial_l^2C \, \partial_i\Phi
     + \left( -\frac{v_0^3(\chi - 20 g v_0)}{1080 D_r^3} + \frac{v_0^2\nu_1}{12 D_r^2} + \frac{5 v_0 \nu_3}{108 D_r^2} \right) 
     \partial_l\partial_iC \, \partial_l\Phi \\
     & - \left( \frac{v_0^2(31\chi^2 + 110\chi g v_0 + 88 g^2 v_0^2)}{1080 D_r^3}+ \frac{v_0\nu_1(\chi+gv_0)}{6 D_r^2} + \frac{v_0\nu_3(20 \chi+23 gv_0)}{270 D_r^2} + \frac{\nu_3^2}{270} \right) 
     \partial_lC \, \partial_l\Phi \, \partial_i\Phi \\
     &- \left( \frac{v_0^2 D}{12 D_r^2} + \frac{v_0^4}{270 D_r^3}\right) \partial_i\partial_l^2C+\mathcal{O}(\nabla^5).
    \end{split}
\end{equation}
\twocolumngrid
Note that at this (mean-field) level, we have not made any assumptions about the chemical field $\Phi$ and therefore the results remain general. The above expression can be used as a basis for constructing the appropriate stochastic field theory description of the system. 
When treating $\Phi$ as the self-generated chemical field, the calculation reveals that there are new chemotactic terms that can play a significant role in determining the collective behavior of such a system.

Note that by setting $v=0$ and $\nu_3=0$, the particle current reduces to the simple form given by Eq.~\eqref{eq:J_final}.

\bibliography{biblio-PRR}

\end{document}